\documentstyle[12pt,epsfig]{article} 
\input epsf
\begin{document}
% This defines greater than, less than, greater or less, less or greater 
% approx. symbols
%
\def\lta{\;\raisebox{-.5ex}{\rlap{$\sim$}} \raisebox{.5ex}{$<$}\;}
\def\gta{\;\raisebox{-.5ex}{\rlap{$\sim$}} \raisebox{.5ex}{$>$}\;}
\def\grle{\;\raisebox{-.5ex}{\rlap{$<$}} \raisebox{.5ex}{$>$}\;}
\def\legr{\;\raisebox{-.5ex}{\rlap{$>$}} \raisebox{.5ex}{$<$}\;}

\newcommand{\ra}{\rightarrow} 
\newcommand{\permille}{$^0 \!\!\!\: / \!_{00}$} 
\newcommand{\dd}{{\rm d}} 
\newcommand{\oal}{\O(\alpha)}%
\newcommand{\su}{$ SU(2) \times U(1)\,$}
 
\newcommand{\np}{Nucl.\,Phys.\,} 
\newcommand{\pl}{Phys.\,Lett.\,}
\newcommand{\pr}{Phys.\,Rev.\,}
\newcommand{\prl}{Phys.\,Rev.\,Lett.\,}
\newcommand{\prep}{Phys.\,Rep.\,} 
\newcommand{\zp}{Z.\,Phys.\,}
\newcommand{\sovjnp}{{\em Sov.\ J.\ Nucl.\ Phys.\ }}
\newcommand{\nuclinst}{{\em Nucl.\ Instrum.\ Meth.\ }}
\newcommand{\annp}{{\em Ann.\ Phys.\ }} 
\newcommand{\intjmp}{{\em Int.\ J.\ of Mod.\ Phys.\ }}

\renewcommand{\O}{{\cal O}} \newcommand{\GeV}{{\rm GeV}}
\newcommand{\eps}{\epsilon} \newcommand{\mw}{M_{W}}
\newcommand{\mt}{m_t} \newcommand{\mb}{m_b}
\newcommand{\mww}{M_{W}^{2}} \newcommand{\mbb}{m_{b \bar b}}
\newcommand{\mcc}{m_{c \bar c}} \newcommand{\mh}{m_{H}}
\newcommand{\mhh}{m_{H}^2} \newcommand{\mz}{M_{Z}}
\newcommand{\mzz}{M_{Z}^{2}} \newcommand{\LamQCD}{\Lambda_{QCD}}
\newcommand{\tgb}{\tan{\beta}} \newcommand{\amu}{a_{\mu}}
\newcommand{\BR}{{\rm BR}} \newcommand{\ms}{{\overline{\rm MS}}}
\newcommand{\lra}{\leftrightarrow} \newcommand{\tr}{{\rm Tr}}
 
\newcommand{\ie}{{\em i.e.}}  
\newcommand{\cm}{{{\cal M}}}
\newcommand{\cl}{{{\cal L}}} 
\def\Ww{{\mbox{\boldmath $W$}}}
\def\B{{\mbox{\boldmath $B$}}} 
\def\nn{\noindent}

\newcommand{\sinsq}{\sin^2\theta} 
\newcommand{\cossq}{\cos^2\theta}
%******

\newcommand{\beq}{\begin{equation}} 
\newcommand{\eeq}{\end{equation}}
\newcommand{\bea}{\begin{eqnarray}} 
\newcommand{\eea}{\end{eqnarray}}

\newcommand{\nl}{\nonumber \\} 
\newcommand{\eqn}[1]{Eq.(\ref{#1})}
\newcommand{\ibidem}{{\it ibidem\/},} 
\newcommand{\into}{\;\;\to\;\;}
\newcommand{\wws}[2]{\langle #1 #2\rangle^{\star}}
\newcommand{\smod}{\tilde{\sigma}}
\newcommand{\dilog}[1]{\mbox{Li}_2\left(#1\right)}
\newcommand{\umu}{^{\mu}} 
\newcommand{\cjg}{^{\star}}
\newcommand{\lgn}[1]{\log\left(#1\right)} 
\newcommand{\si}{\sigma}
\newcommand{\sit}{\sigma_{tot}} 
\newcommand{\sqs}{\sqrt{s}}
\newcommand{\sih}{\hat{\sigma}} 
\newcommand{\sith}{\hat{\sigma}_{tot}}
\newcommand{\p}[1]{{\scriptstyle{\,(#1)}}} 
\newcommand{\res}[3]{$#1\pm #2~~\,10^{-#3}$}
\newcommand{\rrs}[2]{\multicolumn{1}{l|}{$~~~.#1~~10^{#2}$}}
\newcommand{\err}[1]{\multicolumn{1}{l|}{$~~~.#1$}}
\newcommand{\ru}[1]{\raisebox{-.2ex}{#1}}
%*******************
\newcommand{\BXcenu}{B\rightarrow X_c e \nu}
\newcommand{\BXsg}{B\rightarrow X_s \gamma} 
\newcommand{\BXsll}{B\rightarrow X_s \ell^+\ell^-} 
\newcommand{\bXsg}{b\rightarrow X_s \gamma} 
\newcommand{\bsg}{b \rightarrow s \gamma}
\newcommand{\bsglu}{b\rightarrow s g} 
\newcommand{\bsgglu}{b\rightarrow s \gamma g} 
\newcommand{\bsll}{b \rightarrow s \ell^{+}\ell^{-}} 
\newcommand{\bsee}{b \rightarrow s e^{+} e^{-}}
\newcommand{\bsmumu}{b \rightarrow s \mu^{+} \mu^{-}}
\newcommand{\bstautau}{b \rightarrow s \tau^{+} \tau^{-}}
\newcommand{\BRg}{\BR_{\gamma}} 

\newcommand{\BRll}{\BR_{\ell\ell}}
\newcommand{\BRee}{\BR_{ee}}
\newcommand{\BRmm}{\BR_{\mu\mu}}
\newcommand{\BRtt}{\BR_{\tau\tau}}
\newcommand{\All}{{\rm A}_{\ell\ell}}
\newcommand{\Aee}{{\rm A}_{ee}}
\newcommand{\Amm}{{\rm A}_{\mu\mu}}
\newcommand{\Att}{{\rm A}_{\tau\tau}}

\newcommand{\mub}{\mu_b}
\newcommand{\sw}{\sin{\theta_W}} 
\newcommand{\sww}{\sin^2{\theta_W}}
\newcommand{\alphas}{\alpha_s} 
\newcommand{\alphae}{\alpha_{\rm em}}

\def\Frac#1#2{{\displaystyle{#1\over #2}}} \newcommand\xx{\phantom{1}}

\begin{titlepage}
\begin{center}
March 1998
\hfill UND-HEP-97-US03 \\ 
\hfill FTUAM 98/3 \\
\hfill hep-ph/9803451
\vskip .2in
{\large \bf Imminent Phenomenology of a Minimal Gauge-Mediated Model}
\vskip .3in

\vskip .3in
Emidio Gabrielli$^*$\\[.03in]
\vskip 10pt
{\em Departamento de F\'{\i}sica Te\'orica\\ 
Universidad Aut\'onoma de Madrid\\
Cantoblanco, 28049 Madrid\\ Spain}
\vskip 10pt
Uri Sarid$^\dag$\\[.03in]
\vskip 10pt

{\em Department of Physics\\
     University of Notre Dame\\
     Notre Dame, IN 46556\\
     USA}
\end{center}
\vskip 10pt
\begin{abstract}
\medskip
We calculate the inclusive branching ratio for $B \to X_s\gamma$, the inclusive branching ratios and asymmetries for $B \to X_s\ell^+\ell^-$, and the anomalous magnetic moment $g_\mu-2$ of the muon, within a minimal gauge-mediated SUSY-breaking model which naturally generates a large ratio $\tan\beta$ of Higgs field vacuum expectation values. These predictions are highly correlated with each other, depending on only two fundamental parameters: the superpartner mass scale and the logarithm of a common messenger mass. The predictions for $\BXsg$ decay and $g_\mu-2$ are in somewhat better agreement with current experiments than the standard model, but a much sharper comparison will soon be possible using new measurements now in progress or under analysis. 
Moreover we predict large deviations in $B \to X_s e^+ e^-$  and $B \to X_s\mu^+\mu^-$ asymmetries, and somewhat smaller ones in $B \to X_s e^+ e^-$  and $B \to X_s\tau^+\tau^-$ branching ratios, which will be detectable in hadronic colliders.

\end{abstract}
\vskip 60pt
{\small 
 *Electronic address: emidio.gabrielli@roma1.infn.it,~~egabriel@delta.ft.uam.es
\\
\dag Electronic address: sarid@particle.phys.nd.edu
}
\end{titlepage}

\newpage
\section{Introduction}

Arguably the least-understood aspects of supersymmetry (SUSY) as a
physically-acceptable theory are the mechanisms which break SUSY and
which communicate that breaking to the observable sector.  These mechanisms must generate soft-breaking mass terms which
are small enough to protect the Higgs mass, large enough to have
evaded all direct searches so far, and flavor-symmetric enough to
respect all current indirect bounds. In particular, flavour-changing
neutral current (FCNC) processes impose strong constraints on any
non-universal soft-breaking masses, due to gluino- and
photino-mediated SUSY contributions \cite{ref:ggms}.  Moreover
CP-violating phases in the soft-breaking sector must be strongly
suppressed to satisfy the constraints on electric dipole moments
\cite{ref:ggms}.

In supergravity-mediated scenarios with flat K\"ahler 
metric \cite{ref:sugra} the SUSY breaking is communicated
from the hidden sector to the observable one by gravitational
interactions, leading to universality in the soft-breaking 
sector near the Planck scale. But even such a strong assumption about physics at extremely high scales may be not sufficient to prevent large flavor violations at observable scales. Indeed in grand-unified supersymmetric theories \cite{ref:bhs}, the large top Yukawa couplings and radiative effects above the unification scale can generate unacceptably large FCNC and CP-violating processes. Moreover it is doubtful whether supergravity theories derived from superstrings would have flat K\"ahler metrics and flavour-independent supersymmetry-breaking terms \cite{ref:susyuniv}.

Recently there has been a revival of interest in a class of models
which can solve both the SUSY flavour and CP problems: theories of
gauge-mediated SUSY breaking \cite{ref:GM}-\cite{ref:revgr}. In these
theories the SUSY-breaking sector is coupled to the observable
(minimal supersymmetric standard model, or MSSM) sector via gauge
interactions. Many models employ the standard-model (SM) gauge interactions to couple the MSSM to a messenger sector, which then couples more directly to the SUSY-breaking sector (much as in the gravitationally-mediated picture the MSSM and the SUSY-breaking hidden sector communicated only via Planck-scale physics). The messenger sector may be far below the Planck scale, thus also lowering the SUSY-breaking scale, and the gauge interactions which generate observable SUSY-breaking masses are naturally flavor-blind, thus preserving the approximate flavor symmetries of the SM. In the class of models studied in this work, the ultimate source of SUSY breaking is parametrized by a superfield $\hat S$ whose scalar and F components both acquire vacuum expectation values (VEVs) through an unspecified mechanism. Since this superfield is a SM gauge singlet with no direct couplings to the
observable sector, the mechanism of SUSY breaking is largely isolated
from, and therefore unconstrained by, the known properties of the
SM. $\hat S$ does couple to a set of messenger fields, which do carry SM gauge quantum numbers. When their superpartners are split by the couplings to $\hat S$, they radiatively split the MSSM gauginos from the gauge bosons and the sfermions from the SM fermions, thus conveying SUSY breaking to the particles we observe.

The pattern of SUSY breaking in the MSSM produced by gauge mediation
is thus exactly what is needed to solve the SUSY flavor and CP
problems. Moreover the most minimal of these models are highly
predictive, since most of the MSSM parameters beyond the SM depend on only the scale of the messenger masses and of the
SUSY-breaking VEV (and the number of messenger fields). Unfortunately
neither the superpotential $\mu$ term, which is the coupling between
the up- and down-type Higgs superfields, nor the soft-breaking term $B
\mu$, which couples the scalar Higgs fields, is generated by the
usual gauge mediation mechanism: both couplings violate the
Peccei-Quinn (PQ) symmetry of relative rotations between the Higgs
doublets, which gauge interactions preserve. But for phenomenological
reasons $\mu$ must be generated somehow, and indeed a few dynamical
mechanisms
have been proposed to break the PQ symmetry at the proper
scale \cite{ref:dgp},\cite{ref:revgr}. Possible solutions of this $\mu$
problem are an active area of research in this field.

We will regard the $\mu$ term as a free parameter, to be fixed
phenomenologically by requiring that the electroweak vacuum be broken
at the correct scale by radiative corrections. We will then assume, as
in ref.~\cite{ref:rs}, that when the $\mu$ term is generated the
corresponding $B \mu$ term is not (and also the soft-breaking Higgs
masses are unaltered). Of course $B \mu$ will be induced radiatively,
but since we assume that its boundary value at the messenger scale
vanishes, we can predict its value at lower, observable scales
\cite{ref:rs}. The vanishing of $B\mu$ at the messenger scale offers
several attractions:
\begin{enumerate}
\item The two physical SUSY CP violating phases $\arg(B M_{1/2}^*)$
and $\arg(A M_{1/2}^*)$, where $M_{1/2}$ is a gaugino mass and $A$ is
the scalar analog of the Yukawa couplings in the superpotential, are
zero \cite{ref:dns}, solving completely the SUSY CP problem;
\item all soft-breaking terms and relative signs can be predicted
essentially in terms of only two parameters, a common messenger mass
and the SUSY-breaking scale;
\item Since $B\mu$ at observable scales turns out to be small,
$\tan{\beta}$, the ratio between the up- and down-type Higgs VEVs, is
naturally predicted to be large, thus generating the observed large
hierarchy between top and bottom quark masses without appealing to
hierarchical Yukawa couplings.
\end{enumerate}

Since a naturally large $\tgb$ is a signature of this model, we
analyse a class of low energy processes which are particularly
sensitive to large $\tgb$. It is well known that in the MSSM the
processes mediated by magnetic dipole transitions can be enhanced by
$\tgb$. The corresponding SM amplitudes are chiral suppressed, while
some of the SUSY contributions can receive a $\tgb$ enhancement,
making the new contributions competitive with the standard
ones. Within this class of processes, we consider in the framework of
the MGM model the radiative $b\rightarrow s \gamma$ and semileptonic
$b\rightarrow s l^+l^-$ decays (with $l=e,\mu,\tau$) and the
contribution to the anomalous magnetic moment of the muon
$g_{\mu}-2$. For the $\bsll$ processes we analyse both the decay
rate and the energy or forward-backward (FB) asymmetries.  

In $b\rightarrow s \gamma$ and $g_{\mu}-2$ the
corresponding amplitudes are proportional to the magnetic dipole
transitions and so are directly enhanced by $\tgb$. The
$b\rightarrow s l^+l^-$ amplitude receives contributions from both the local four-quark operators $Q_9$ and $Q_{10}$, which are not enhanced by $\tgb$ and in fact will be largely unaffected by the MGM, and from the same magnetic-dipole operator $Q_7$ responsible for the $\bsg$ decay. ($Q_7$ contributes to the $\bsll$ amplitude through the exchange diagram between $Q_7$ and the electromagnetic lepton current.) Therefore the sensitivity of $\bsll$ to $\tgb$ is correlated with that of $\bsg$ through their common dependence on a single Wilson coefficient.

A number of studies for $\bsg$ \cite{ref:bbmr}--\cite{ref:bsg2}, $\bsll$ \cite{ref:bbmr},\cite{ref:bsllsusy}--\cite{ref:hewett} and $g_{\mu}-2$ \cite{ref:gmtwoSUSYgen}--\cite{ref:cgw} have already been carried out for MSSM models similar to the one we consider. But our analysis for the MGM model is unique in its predictivity: there are essentially only two parameters, and they determine a tight correlation between these processes. (Moreover in our analysis we include the complete next-to-leading order accuracy in the strong coupling for the SM contribution to $b\rightarrow s \gamma$ \cite{ref:bsgNLO1} and $b\rightarrow s l^+l^-$ \cite{ref:bsll1},\cite{ref:bsll2}.) Present data {\it mildly} favor our predictions over those of the SM. Measuring the correlated deviations in these processes from SM predictions will either favor this model over the standard one and measure its parameters, or strongly constrain the model by forcing it to agree closely with the SM, or disfavor both this model and the standard one.  

The paper is organized as follows : in the next section we define the model and give the analytical results for the mass spectrum of squarks, sleptons, Higgs and gauginos at the physical scale. We state these results for the mildly more general model than the minimal one, allowing for $N > 1$ messenger families. All these quantities are expressed as a function of the 
SUSY-breaking scale $\Lambda$ (or equivalently the wino mass) and the common messenger mass. Also in this section we give the numerical predictions for the 
magnitude of the $\mu$ parameter (for various values of N, messenger scale and top mass) and a detailed discussion about the predictions of its sign.
In Secs. 3, 4 and 5 we analyse the contribution of the MGM model
to the $\bsg$, $\bsll$ decay rates and asymmetries, and $g_\mu-2$, respectively. We give model-independent parametrizations for the total rates and asymmetries and analytical expressions for the MGM contributions to 
the corresponding amplitudes in a suitable approximation. In Sec. 6 
we present and analyse the numerical predictions for the $\bsg$ and $\bsll$ decay rates and asymmetries versus the $g_\mu-2$ ones, as a function of the SUSY breaking scale and the common messenger mass in both $N=1$ and $N=2$ scenarios. 
The final section contains our conclusions and outlook.

\section{Minimal gauge mediated model}
In GMSB models supersymmetry is broken at some scale above the
electroweak scale and then the breaking is communicated to the MSSM
particles by the usual SM gauge 
interactions~\cite{ref:GM}-\cite{ref:revgr}.  In the
minimal realization of this idea, a pair of SU(2) doublet chiral
superfields $\Phi_2$, $\overline\Phi_2$ and a pair of SU(3) triplet
chiral superfields $\Phi_3$, $\overline\Phi_3$, called messenger
fields, couple to a singlet chiral superfield $\hat{S}$ with Yukawa
couplings $\lambda_{2,3}$.  The messenger fields get their
supersymmetric mass through the VEV $\langle S \rangle$ of the scalar
component of the $\hat{S}$ superfield.  But the $F$ component of
$\hat{S}$ also has a nonvanishing VEV, $\langle F \rangle$, which
parametrizes the breaking of supersymmetry. The breaking is manifest
in the consequent mass splitting between the fermionic and scalar
component of the messenger fields $\Phi_i$: the fermions of $\Phi_i$
acquire masses $M_{M_i}=\lambda_i \langle S \rangle$ while their
scalar partners get squared masses $m^2_i=|\lambda_i\langle S \rangle
|^2\pm|\lambda_i \langle F \rangle |$.

In this minimal gauge-mediation (MGM) model, the messenger fields
couple at tree level only with the gauge sector of the MSSM and not
with its matter fields. Thus SUSY breaking is transmitted to the MSSM
only by the ordinary gauge interactions. At one loop, gaugino masses
are generated by finite self-energy diagrams, with the fermionic and
scalar components of the messenger fields running in the loop. In the
following we will allow the possibility of $N \geq 1$ messenger
families (pairs of doublets and triplets), but only a single source
$\hat S$ of SUSY breaking. We will also assume that the messengers are
approximately degenerate; our results are not very sensitive to this
splitting as long as the messengers are not very split and not very
close to their lower bound (see below). We denote quantities evaluated
at the messenger scale by an overbar. Then the gaugino mass spectrum
$\overline{M}_i$ is given by
\beq \overline{M}_i = N \frac{\overline{\alpha_i}}{4\pi}\Lambda g(x_i)
\equiv \hat{M}_i g(x_i)
\label{eq:gauginomass}
\eeq
where $\Lambda \equiv \langle F_S \rangle/ \langle S \rangle$ and $x_i
= \Lambda/M_{M_i}$. The function $g(x)$ contains the result of the
one-loop integration.  Its argument must satisfy $x_i < 1$ to ensure
that the messenger scalars have a positive mass-squared. We will not
be concerned with $M_{M_i}$ just above but very close to the minimal
value $\Lambda$, so $x_i$ will never approach unity, in which case
$g(x_i)\simeq 1$ and $\overline{M}_i \simeq \hat M_i$ will be very
good approximations.

At two-loop order in the messenger fields, or more plainly at one-loop
order in the gaugino masses, squark, slepton and Higgs SUSY-breaking
{\it squared} masses are generated by self-energy diagrams with a
gaugino mass insertion. Thus the sfermion masses (rather than their
squares) are the same order as the gaugino masses. Once again we are
concerned with $x_i$ well below unity, in which case the results of
the sfermion mass calculation can be expressed as~\cite{ref:rs}:
\beq \overline{m}^2_{\alpha} \simeq \frac1N\left(2 C_3 \hat M_3^2 + 2
C_2 \hat M_2^2+\frac65 Y^2 \hat M_1^2\right)
\label{eq:sfermionmass}
\eeq
where the coefficients $C_i$ depend on the gauge quantum numbers of
the scalar multiplets and $Y$ is the weak hypercharge. Explicitly,
these equations read:
\bea \overline{m}^2_{\tilde{Q}}&=&{\bf
1}\,\frac1N\left(\frac{8}{3}\hat{M}_3^2+\frac{3}{2}\hat{M}^2_2+
\frac{1}{30}\hat{M}_1^2\right) 
\label{eq:squarksleptonmass1}
\\ \overline{m}^2_{\tilde{u}}&=&{\bf 1}
\,\frac1N\left(\frac{8}{3}\hat{M}_3^2+\frac{8}{15}\hat{M}_1^2
\right)\\ \overline{m}^2_{\tilde{d}}&=&{\bf 1}
\,\frac1N\left(\frac{8}{3}\hat{M}_3^2+\frac{2}{15}\hat{M}_1^2
\right)\\
\overline{m}^2_{\tilde{L}}&=&{\bf
1}\,\frac1N\left(\frac{3}{2}\hat{M}_2^2+
\frac{3}{10}\hat{M}^2_1\right) \\ \overline{m}^2_{\tilde{l}}&=&{\bf 1}
\,\frac1N\left(\frac{6}{5}\hat{M}_1^2 \right)\\
\overline{m}^2_{H_U}=\overline{m}^2_{H_D}&=&\phantom{{\bf
1}}\frac1N\left(\frac{3}{2}\hat{M}^2_2+\frac{3}{10}\hat{M}^2_1\right)
\label{eq:squarksleptonmass2}
\eea
for the squared masses of the squark doublet, the up- and down-type
squark singlets, the slepton doublet, the charged slepton singlet, and
the up- and down-type Higgs doublets. We've also explicitly inserted
the unit matrix ${\bf 1}$ in flavor space: 
the gauge interactions which generate these masses are flavor-blind.

Another consequence of the pure gauge mediation in the MGM is that the
SUSY-breaking $A$ terms which couple the squarks (or sleptons) to the
Higgs doublets are not induced at the same order as the gaugino or
sfermion masses, since they break the chiral (flavor) symmetries while
the gauge interactions do not. $A$ terms are induced at higher order
via gaugino masses, so both $A$ and $\hat M$ arise through the same
source $\Lambda$, and hence the relative CP-violating phase between
them $\mbox{arg}(A_i^* M_3)$ vanishes naturally.

The $\mu$ parameter, which couples the up- and down-type Higgs doublet
superfields in the superpotential, is not generated within the GMSB
paradigm, since it breaks the Peccei-Quinn symmetry while the gauge
interactions do not. To have a (phenomenologically-mandated)
nonvanishing $\mu$ term, the MGM model must be extended. We will {\it
assume} that such an extension does not have any impact on the mass
parameters beyond the $\mu$ term (but see Ref.~\cite{ref:dgp} for
counterexamples). Thus effectively $\mu$ is an arbitrary additional
parameter of the model, as was assumed in most of the previous
literature.  But in fact its value can be predicted by the
phenomenological requirement that the electroweak vacuum should
spontaneously break to yield the proper Higgs VEV.

The SUSY-breaking analogue of the $\mu$ term, the $B\mu$ term coupling
the Higgs doublets in the scalar potential, is also not generated by
the pure MGM. We will {\it assume}, as we did above, that whatever
mechanism generates $\mu$ does {\it not} generate $B\mu$, and this is
the major assumption of this work. While this is by no means a generic
aspect of GMSB models (see Ref.~\cite{ref:dgp} for critical analysis of
this issue), it is theoretically conceivable since $B\mu$ would break
an R symmetry which $\mu$ would not (see Ref.~\cite{ref:dns} 
for an example and Ref.~\cite{ref:rs} for further discussion). It is also
phenomenologically appealing, since (as for the $A$ terms) $B$ is
generated radiatively through gaugino masses, and hence the
CP-violating relative phase $\mbox{arg}(B^* M_3)$ once again vanishes
naturally. Thus the SUSY CP problem is completely absent in this
model. To reiterate: in the following we will always assume the
boundary condition $\overline{B}\simeq 0$.

Notice that the only source of phases (apart from topological vacuum
angles) is the SM CP-violating phase in the quark Yukawa couplings. In
particular, no physical relative signs are arbitrary parameters; thus
the relative sign between $\mu$, the Higgs VEVs and the gaugino masses
(which is often just called the sign of $\mu$) is a {\it prediction}
of this model, which we calculate below.

Since the messenger scale is considerably higher than the electroweak
scale, we use the above MSSM parameters as boundary conditions to the
renormalization group (RG) evolution equations, and calculate the
RG-improved masses closer to the electroweak scale. But as long as the
messenger scale is only a few orders of magnitude above the electroweak
scale, a one-step solution to the RG equations will suffice for all
parameters except $B$, for which a delicate cancellation occurs. The
RG evolution starts at the messenger scale; it effectively stops at a
typical squark mass scale $\tilde m$. Both of these statements were
discussed and justified in Ref.~\cite{ref:rs}; in particular, the
doublet and triplet messenger masses are expected to be quite similar
in a unified model, so there is a reasonably well-defined single
messenger scale, and most of the results do not depend significantly
on the small splitting between $M_{M_2}$ and $M_{M_3}$. We will
usually only keep the leading powers of $L/(8\pi^2) = \ln(M_M/\tilde
m)/(8\pi^2)$, where $M_{M}$ is the messenger mass scale. Since
$\tan\beta= \langle H_U\rangle / \langle H_D\rangle$ will also turn
out large, we will also usually keep only leading powers of
$1/\tan\beta$.

The gaugino masses, and thence the physical chargino and neutralino
mass matrices, are given at the squark scale respectively by: \bea
M_i&=&\left(\frac{\alpha_i}{\overline{\alpha}_i}\right) \hat{M}_i
\label{eq:gauginomassLOW}
\\ M_{\chi^{\pm}}&=&\left( \matrix{ M_2 & \sqrt{2}\mw \cr 0 & \mu
\cr}\right)
\label{eq:charginomass}
\\ M_{\chi^{0}}&=&\left( \matrix{ M_1 & 0 & 0 & \sw m_Z \cr 0 & M_2 &
0 & -\cos{\theta_{W}} m_Z \cr 0 & 0 & 0 & -\mu \cr \sw m_Z &
-\cos{\theta_{W}} m_Z & -\mu & 0 \cr }\right)
\label{eq:neutralinomass}
\eea

The up- and down-type squark mass matrices $M^2_{\tilde{u}}$ and
$M^2_{\tilde{d}}$ respectively at the squark scale are
\beq M^2_{\tilde{u}}=\left( \matrix{ m^2_{\tilde{Q}}+m^2_{u}-{\bf
1}\left(\frac{1}{2}-\frac{2}{3}\sww\right)m_Z^2 & -m_{u} A_U\cr -m_{u}
A_U & m^2_{\tilde{u}}+m^2_{u}-{\bf 1}\frac{2}{3}\sww m_Z^2\cr} \right)
.
\label{eq:squarkmassU}
\eeq
\beq M^2_{\tilde{d}}=\left( \matrix{ m^2_{\tilde{Q}}+m^2_{d}+{\bf
1}\left(\frac{1}{2}-\frac{1}{3}\sww\right)m_Z^2 & \mu m_{d}\tgb \cr
\mu m_{d}\tgb & m^2_{\tilde{d}}+m^2_{d}+{\bf 1}\frac{1}{3}\sww
m_Z^2\cr} \right) .
\label{eq:squarkmassD}
\eeq where \bea
m^2_{\tilde{Q}}&=&\overline{m}^2_{\tilde{Q}}+\left\{{\bf 1}
\left(\frac{16}{3}g_3^2 \overline{M}_3^2+ 3 g_2^2 \overline{M}_2^2
+\frac{1}{15}g_1^2 \overline{M}_1^2 \right) -{\lambda^2_{U}\over
N}\left(\frac{16}{3} \hat{M}_3^2+ 3 \hat{M}_2^2
+\frac{13}{15}\hat{M}_1^2\right) \right. \nl &-& \left.
{\lambda^2_{D}\over N}\left(\frac{16}{3} \hat{M}_3^2+ 3 \hat{M}_2^2
+\frac{7}{15}\hat{M}_1^2\right)\right\} {L\over8\pi^2} \eea
\bea m^2_{\tilde{u}}&=&\overline{m}^2_{\tilde{u}}+\left\{{\bf 1}
\left(\frac{16}{3}g_3^2 \overline{M}_3^2+ \frac{16}{15}g_1^2
\overline{M}_1^2 \right) -{2\lambda^2_{U}\over N}\left(\frac{16}{3}
\hat{M}_3^2+ 3 \hat{M}_2^2 +\frac{13}{15}\hat{M}_1^2\right) \right\}
{L\over8\pi^2} \eea
\bea m^2_{\tilde{d}}&=&\overline{m}^2_{\tilde{d}}+\left\{{\bf 1}
\left(\frac{16}{3}g_3^2 \overline{M}_3^2+ \frac{4}{15}g_1^2
\overline{M}_1^2 \right) -{2\lambda^2_{U}\over N}\left(\frac{16}{3}
\hat{M}_3^2+ 3 \hat{M}_2^2 +\frac{7}{15}\hat{M}_1^2\right) \right\}
{L\over8\pi^2}.  \eea
$m_u~(\lambda_U)$ and $m_d~(\lambda_D)$ are the up- and down-type
quark mass (Yukawa coupling) matrices in flavour space, and $A_U$ is
given by \beq A_U={\bf
1}\left(\frac{16}{3}g_3^2\overline{M}_3+3g_2^2\overline{M}_2+
\frac{13}{15}g_1^2\overline{M}_1\right) {L\over8\pi^2}
\label{eq:Aterm}
\eeq where the boundary condition $\overline{A}_U\simeq 0$ has been
used.

For the slepton mass matrices we find, with the same approximations as
for the squarks, 
\bea 
M^2_{\tilde{L}}&=&\left( \matrix{
m^2_{\tilde{L}}+m^2_{l}+{\bf 1}\left(\frac{1}{2}-\sww\right)m_Z^2 &
\mu m_l\tgb\cr \mu m_l\tgb & m^2_{\tilde{l}}+{\bf 1}\sww m_Z^2\cr}
\right) \label{eq:sleptonmass} \\
M^2_{\tilde{\nu}}&=&\left(m^2_{\tilde{L}}-{\bf 1}
\frac{1}{2}m_Z^2\right)
\label{eq:sneumass}
\eea where \bea
m^2_{\tilde{L}}&=&\overline{m}^2_{\tilde{L}}+\left\{{\bf 1} \left(3
g_2^2 \overline{M}_2^2 +\frac{3}{5}g_1^2 \overline{M}_1^2 \right)
-{\lambda^2_{l}\over
N}\left(3\hat{M}_2^2+\frac{9}{5}\hat{M}_1^2\right) \right\}
{L\over8\pi^2} \\
m^2_{\tilde{l}}&=& \overline{m}^2_{\tilde{l}}+\left\{{\bf 1}
\left(\frac{12}{5}g_1^2 \overline{M}_1^2\right) -{2\lambda^2_{l}\over
N}\left(3 \hat{M}_2^2 +\frac{9}{5}\hat{M}_1^2\right) \right\}
{L\over8\pi^2} \eea
and $m_\l$ ($\lambda_\l$) are the diagonal lepton mass (Yukawa) matrix
in lepton flavour space.

The remaining parameters to be determined are the Higgs couplings
$\mu$ and $B \mu$. These enter the scalar potential and determine the
VEVs of the two Higgs doublets $v_U$ and $v_D$ in terms of the
soft-breaking masses (discussed above) and the quartic Higgs couplings
(fixed supersymmetrically by the gauge couplings). Since we are
assuming $\overline B = 0$, the only free parameter is $\mu$, and it
is set by requiring agreement with the experimental observable $v^2 =
v_U^2 + v_D^2 \simeq (174\,\GeV)^2$, or equivalently with $m_Z^2 =
(g_1^2 + g_2^2) v^2/2$. Since the coupling $B\mu$ between the up- and
down-type Higgs scalars will be very small, $v \simeq v_U \gg v_D$, so
the electroweak scale requirement is essentially a condition on the
up-type Higgs mass-squared: $m_{H_U}^2 = -\Frac12 m_Z^2$ to leading
order. Substituting in the up-type Higgs mass as a function of
$\mu^2$, and including higher-order corrections, leads to:
\beq \mu^2 = \lambda_t^2 \Delta_t^2 - m_H^2 -\frac12 m_Z^2 (1 +
\delta_H).
\label{eq:mueq}
\eeq
The details of this minimization are given in Ref.~\cite{ref:rs}:
$\delta_H$ ($\sim 1$) accounts for the loop corrections to the Higgs
quartic couplings induced by the heavy top, $m_H^2 \simeq \Frac1N
\left(\Frac32 \hat M_2^2 + \Frac3{10} \hat M_1^2\right) + 3 \overline
g_2^2 \overline M_2^2 \left(\Frac{L}{8\pi^2}\right)$ is the common
Higgs mass, and $\lambda_t^2 \Delta_t^2$ is the Yukawa coupling
correction which lowers the up-type Higgs mass and breaks the
electroweak symmetry. Neglecting terms smaller than $\sim1\%$, and
recalling that $\lambda_t$ is evaluated at the squark mass scale as in
Ref.~\cite{ref:rs}, we find:
\beq \Delta_t^2 \simeq {3\over N} {L_\Delta\over 8\pi^2} \left\{
\left[\frac{16}{3} \hat M_3^2 + 3 \hat M_2^2\right] + {L_\Delta\over
8\pi^2} \left[N \frac{16}{3} \overline g_3^2 \overline M_3^2 -
\left(\frac{128}{9} \overline g_3^2 + 8 \overline g_2^2\right) \hat
M_3^2 - 8 \overline g_3^2 \hat M_2^2\right] \right\}
\label{eq:Dtappr}
\eeq
We will set the logarithm to be $L_\Delta = \ln (M_{M_3}/m_{\tilde Q})
+ 3/2$ where the constant term, calculated in Ref.~\cite{ref:gr},
accounts for the 1-loop thresholds at the messenger scale. The absence
of the complete 1-loop threshold expression resulted in the major
uncertainty in the previous analysis of this model \cite{ref:rs}.

Numerically, we find that $\mu$ increases approximately linearly with
the superpartner scale, which we characterize by the wino mass $M_2$,
even for light $M_2$. Specifically, we find that $\mu$ is given by the
linear expressions of Tables~\ref{tab1}--\ref{tab4}
 for various values of $N$, the
messenger scale and the top mass. These approximations yield $\mu$
within $\sim\pm10\,\GeV$ over the range $100\,\GeV < M_2 <
400\,\GeV$. Several features are evident in these results: $\mu$
increases with the messenger scale and with the top mass, since it
must cancel a larger radiative correction $\lambda_t^2 \Delta_t^2$; it
decreases with $N$ relative to $M_2$, since more messengers lower the
sfermions relative to the gauginos [see
Eqs.~(\ref{eq:gauginomass},\ref{eq:sfermionmass})]; and it's rather
insensitive to small messenger splittings induced presumably by RG
evolution from high scales.

\begin{table}
\begin{center}
\begin{tabular}{|r||c|c|c|}
\hline
$M_{M_2}/\Lambda$ & $m_t = 165\,\GeV$ & $m_t = 175\,\GeV$ & $m_t =
185\,\GeV$ \\ \hline \hline 1 & $23\,\GeV + 1.08 M_2$ & $23\,\GeV +
1.22 M_2$ & $23\,\GeV + 1.36 M_2$ \\ \hline 2 & $22\,\GeV + 1.46 M_2$
& $23\,\GeV + 1.64 M_2$ & $23\,\GeV + 1.82 M_2$ \\ \hline 10 &
$22\,\GeV + 1.62 M_2$ & $22\,\GeV + 1.81 M_2$ & $22\,\GeV + 2.01 M_2$
\\ \hline 100 & $20\,\GeV + 1.70 M_2$ & $20\,\GeV + 1.90 M_2$ &
$20\,\GeV + 2.11 M_2$ \\ \hline 1000 & $18\,\GeV + 1.74 M_2$ &
$18\,\GeV + 1.95 M_2$ & $17\,\GeV + 2.16 M_2$ \\ \hline 10000 &
$15\,\GeV + 1.76 M_2$ & $15\,\GeV + 1.97 M_2$ & $14\,\GeV + 2.19 M_2$
\\ \hline
\end{tabular}
\caption[]{The MGM predictions for $\mu$ as a function of the wino mass $M_2$ 
in the linear approximation and for some values of messenger mass $M_{M_2}$
and top mass $m_t$, with $M_{M_3} = 1.3 M_{M_2}$ and $N=1$.}
\label{tab1}
\end{center}
\end{table}

\begin{table}
\begin{center}
\begin{tabular}{|r||c|c|c|}
\hline
$M_{M_3}/\Lambda$ & $m_t = 165\,\GeV$ & $m_t = 175\,\GeV$ & $m_t =
185\,\GeV$ \\ \hline \hline 1 & $22\,\GeV + 1.26 M_2$ & $22\,\GeV +
1.43 M_2$ & $22\,\GeV + 1.60 M_2$ \\ \hline 2 & $21\,\GeV + 1.43 M_2$
& $21\,\GeV + 1.61 M_2$ & $21\,\GeV + 1.80 M_2$ \\ \hline 10 &
$21\,\GeV + 1.57 M_2$ & $21\,\GeV + 1.76 M_2$ & $20\,\GeV + 1.96 M_2$
\\ \hline 100 & $19\,\GeV + 1.66 M_2$ & $19\,\GeV + 1.86 M_2$ &
$19\,\GeV + 2.07 M_2$ \\ \hline 1000 & $17\,\GeV + 1.71 M_2$ &
$17\,\GeV + 1.91 M_2$ & $16\,\GeV + 2.12 M_2$ \\ \hline 10000 &
$15\,\GeV + 1.73 M_2$ & $15\,\GeV + 1.94 M_2$ & $14\,\GeV + 2.15 M_2$
\\ \hline
\end{tabular}
\caption[]{The MGM predictions for $\mu$ as a function of the wino mass $M_2$ 
in the linear approximation and for some values of messenger mass $M_{M_3}$
and top mass $m_t$, with $M_{M_2} = 1.3 M_{M_3}$ and $N=1$.}
\label{tab2}
\end{center}
\end{table}

\begin{table}
\begin{center}
\begin{tabular}{|r||c|c|c|}
\hline
$M_{M_2}/\Lambda$ & $m_t = 165\,\GeV$ & $m_t = 175\,\GeV$ & $m_t =
185\,\GeV$ \\ \hline \hline 1 & $\xx 8\,\GeV + 0.81 M_2$ & $\xx
9\,\GeV + 0.91 M_2$ & $10\,\GeV + 1.01 M_2$ \\ \hline 2 & $\xx 8\,\GeV
+ 1.10 M_2$ & $\xx 9\,\GeV + 1.23 M_2$ & $10\,\GeV + 1.36 M_2$ \\
\hline 10 & $\xx 9\,\GeV + 1.23 M_2$ & $10\,\GeV + 1.37 M_2$ &
$11\,\GeV + 1.51 M_2$ \\ \hline 100 & $\xx 9\,\GeV + 1.31 M_2$ &
$10\,\GeV + 1.46 M_2$ & $10\,\GeV + 1.60 M_2$ \\ \hline 1000 & $\xx
8\,\GeV + 1.35 M_2$ & $\xx 9\,\GeV + 1.50 M_2$ & $\xx 9\,\GeV + 1.65
M_2$ \\ \hline 10000 & $\xx 7\,\GeV + 1.37 M_2$ & $\xx 8\,\GeV + 1.53
M_2$ & $\xx 8\,\GeV + 1.68 M_2$ \\ \hline
\end{tabular}
\caption[]{The $\mu$ MGM predictions as in Table 1 with $N=2$}
\label{tab3}
\end{center}
\end{table}

\begin{table}
\begin{center}
\begin{tabular}{|r||c|c|c|}
\hline
$M_{M_3}/\Lambda$ & $m_t = 165\,\GeV$ & $m_t = 175\,\GeV$ & $m_t =
185\,\GeV$ \\ \hline \hline 1 & $\xx 7\,\GeV + 0.96 M_2$ & $\xx
9\,\GeV + 1.08 M_2$ & $10\,\GeV + 1.20 M_2$ \\ \hline 2 & $\xx 6\,\GeV
+ 1.08 M_2$ & $\xx 7\,\GeV + 1.21 M_2$ & $\xx 8\,\GeV + 1.34 M_2$ \\
\hline 10 & $\xx 7\,\GeV + 1.19 M_2$ & $\xx 9\,\GeV + 1.33 M_2$ & $\xx
9\,\GeV + 1.47 M_2$ \\ \hline 100 & $\xx 8\,\GeV + 1.27 M_2$ & $\xx
9\,\GeV + 1.42 M_2$ & $\xx 9\,\GeV + 1.56 M_2$ \\ \hline 1000 & $\xx
7\,\GeV + 1.32 M_2$ & $\xx 8\,\GeV + 1.47 M_2$ & $\xx 8\,\GeV + 1.62
M_2$ \\ \hline 10000 & $\xx 6\,\GeV + 1.34 M_2$ & $\xx 7\,\GeV + 1.50
M_2$ & $\xx 7\,\GeV + 1.65 M_2$ \\ \hline
\end{tabular}
\caption[]{The $\mu$ MGM predictions as in Table 2 for $N=2$}
\label{tab4}
\end{center}
\end{table}

Finally, $\tgb = v_U/v_D$ is predicted by minimizing the full scalar
potential for both Higgs doublets:
\beq \frac{1}{\tgb}=-\frac{B\mu}{m^2_{{\rm eff},A}}
\label{eq:tgb}
\eeq
A striking aspect of the MGM is that, assuming $\overline B = 0$ (or
practically zero), the low-energy value of $B$ is also very small, and
hence $\tan\beta$ is naturally very large~\cite{ref:bkw}, \cite{ref:rs}. 
The reason is not simply a small amount of RG evolution
between the messenger and squark mass scales, but rather \cite{ref:rs}
a cancellation between the direct gaugino contribution to the RG
evolution and the direct $A$-term (or indirect gluino) contribution to
that evolution --- a fortuitous cancellation that happens when $M_M$
is not too far above its minimal value $\Lambda$. Thus if the
messengers are not too heavy, the $B$ parameter evaluated at the
squark mass scale is very small, as it is at the messenger scale,
although for intermediate scales $B$ is considerably larger. So the
coupling between the Higgs scalars is predicted to be very small at
the relevant scale of electroweak breaking, which leads to a large
hierarchy between their VEVs: $\tgb = v_U/v_D \gg 1$. Because of the
cancellation of the leading-order terms, the $B(m_{\tilde Q})$ must be
calculated quite carefully at the next-to-leading order. The analytic
results of this calculation, and the expression for the effective
pseudoscalar mass-squared parameter $m^2_{{\rm eff},A}$ to be used in
minimizing the potential, are given in Ref.~\cite{ref:rs}. Then
$\tan\beta$ is predicted by solving Eq.~(\ref{eq:tgb}) numerically,
taking into account the implicit dependence of the right-hand side
(and in particular of $B$) on $\tan\beta$.

In addition to new mass scale parameters, the MSSM is characterized by
two new phase parameters beyond the SM. The MGM predicts these signs in any fixed convention. To see this, note that they may be put
entirely into the $A$ and $B$ parameters by appropriate field
redefinitions \cite{ref:gnr}. But in minimal gauge mediation the $A$ and $B$
parameters of the effective MSSM at the messenger scale vanish, and
hence the phase parameters are predicted: they vanish. Then their
values at lower scales follow from the calculable RG evolution of $A$
and $B$. In a convention where the gaugino masses and top and bottom Yukawa couplings are real, no complex phases can then be generated (at least in low orders) in $A$ and $B$, only a relative sign, which we calculate below. 

This sign is particularly crucial for the observables of interest to us, $g_\mu-2$ and $b\to s\gamma$, which are enhanced by large $\tgb$. As shown in
the first reference of \cite{ref:bsg1}, all such observables proportional 
to $\tgb$ are also proportional to the degree of breaking of the 
Peccei-Quinn (PQ) symmetry of the Higgs sector, characterized by $\mu$, 
and of an R symmetry, characterized by the gaugino mass $M_{1/2}$. 
The amplitudes of these observables will be determined by the sign of the product $\tgb \mu M_{1/2}$, which is a prediction of the MGM in any fixed Lagrangian convention, as we calculate below. The final predictions of the observables will of course be independent of any convention.

To state the sign, we must establish some convention. In 2-component Weyl notation, we use a Lagrangian schematically of the form:
\bea {\cal L}_{\rm gauge} &\sim& i g \left(\widetilde G t_L \tilde t_L
- \widetilde G t_R \tilde t_R\right)
\label{eq:LG}\\
{\cal L}_{\rm Yukawa} &\sim& -\int d^2\theta W,\quad W \sim \mu H_U^0
H_D^0 + \lambda_t H_U^0 t_L t_R
\label{eq:LY}\\
{\cal L}_{\rm soft} &\sim& - M_{1/2} \widetilde G \widetilde G
-\lambda_t A_t \tilde t_L \tilde t_R H_U^0 -B \mu H_U^0 H_D^0
\label{eq:LS}\\
{\cal L}_{\rm F} &\sim& - \left|\partial W/\partial \phi\right|^2 \sim
- \lambda_t \mu H_D^0 \tilde t_L \tilde t_R
\label{eq:LF}
\eea
Here $H_{U,D}^0$ are the neutral components of the Higgs doublets,
$\widetilde G$ is the gaugino field, and the proper group structure
and indices are implied: our purpose is only to establish the sign
convention. With this Lagrangian, we calculate (and find agreement
with Moroi \cite{ref:moroi}) that, for example, the dominant MSSM contribution to $g_\mu - 2$ is $\Delta a \sim + M_{1/2} \mu \tgb$. We can also easily
extract and minimize the scalar Higgs potential to find $1/\tgb \sim -
B \mu$. Next, a 1-loop top-gluino diagram using this Lagrangian
generates $A_t \sim +M_{1/2}$. Then $B$ is generated by both a 1-loop
stop diagram with an $A$ insertion, $B_A \sim - A_t \sim -M_{1/2}$,
and a 1-loop higgsino-wino diagram, $B_G \sim +M_{1/2}$. The
coincidental near-cancellation of these two is the reason $\tgb$ is
large in the MGM, but we find that essentially always $|B_A| > |B_G|$,
so $B \sim -M_{1/2}$. Therefore with this sign convention $M_{1/2} \mu
\tgb > 0$. Hence the convention-{\it independent}, {\it physical} sign
of $\Delta a$ is positive, and similarly the deviations of the various
$b\to s \gamma$ amplitudes are unambiguously predicted.

To be more specific, for the remainder of this paper we will further specify that $\mu$ and $M_{1/2}$ are positive, and therefore so is $\tgb$. To translate our intermediate results to different Lagrangian conventions, the interested reader need only use the Lagrangian of Eqs. (\ref{eq:LG}) - (\ref{eq:LF}) to determine the desired signs of $\mu$ and $M_{1/2}$, and then establish the corresponding sign of $\tgb$ from $M_{1/2} \mu \tgb > 0$. Our final results are physical, and therefore will be unchanged. In much of the existing literature, the signs of $M_{1/2}$ and $\tgb$ are fixed by convention, in which case the MGM makes a prediction of the sign of $\mu$.

We conclude this section by observing that the low energy mass spectrum
in the MSSM, as well as $\tan\beta$ and the sign of $\mu$, can be
predicted within the MGM in terms of only the SUSY-breaking scale
$\Lambda$ (or equivalently any particular superpartner mass, say the
wino $M_2$) and the logarithm of the messenger mass scale
$M_{M}$. This extraordinary predictivity forms the basis for the
remainder of this work: the formulation of low-energy experimental
tests of the MGM.

%%%%%%%%%%%%%%%%%%%%%%%%%%%%%%%%%%%%%%%%%%%%%%%%%%%%%%%%%%%%%%%%%%%%%%%%%%%%
%\newpage
\section{${\bf \bsg }$}

As discussed in the previous section, an interesting feature of the MGM
model is that $\tgb$ is predicted large. Since we are interested in
constraining this model from low energy physics, we will restrict
our investigation to some processes which are particularly sensitive
to large $\tgb$ values. It is well known that the low energy processes
mediated by magnetic dipole transitions can be significantly
enhanced in the MSSM for $\tgb \gg1$, because the chiral suppression
of superpartner-mediated diagrams is removed to a large extent by the
large $\tan\beta$ (i.e. the Yukawa couplings are no longer small),
which allows for a competition with the chirally-suppressed SM
amplitude.  Among this class of processes we consider here the rare
$\BXsg$ decay. In particular in this section we will give the
analytical results for the dominant contributions to the $\BXsg$ decay
in the MGM model, and fold them in with the recent Next to Leading
Order (NLO) calculation for the SM.

Let us start with the experimental results. The most recent published
result by the CLEO collaboration for the total {\it inclusive}
branching ratio ${\rm BR}(\BXsg)$ is~\cite{ref:bsgEXPT} 
\beq 
10^4 \BR^{\rm expt}(\BXsg )= 2.32 \pm 0.67
\label{eq:bsgEXP}
\eeq 
combining $1\,\sigma$ uncertainties, and 
\beq 
1.0 < 10^4 \BR^{\rm expt}(\BXsg ) < 4.2 
\label{eq:bsgEXP95}
\eeq 
at the 95\% confidence level.

Recently the NLO corrections to the quark $\bsg$ decay rate have been
completed in Ref.~\cite{ref:bsgNLO1}, including the calculation of the
three-loop anomalous dimension matrix of the effective theory.  This
is a necessary ingredient for a consistent resummation, at the NLO
accuracy, of the large leading logarithms, $\log{\mw/\mb}$.  

The inclusion of the NLO corrections in the $\bsg$ decay rate has
significantly reduced the large theoretical uncertainty present in the
previous LO calculation.  From the quark-level $\bsg$ decay rate it is
possible to infer the B meson {\it inclusive} branching ratio $\BRg
\equiv \BR(\BXsg)$ by including the small (percent-level)
non-perturbative $1/\mb$ \cite{ref:bsgNPmb} and $1/m_c$
\cite{ref:bsgNPmc},~\cite{ref:bsgllNPmc} corrections.  The result,
the most recent theoretical prediction for $\BRg^{\rm NLO}$ in the SM,
is~\cite{ref:bsgNLO2} 
\beq 
10^4 \BRg^{\rm NLO} =3.48 \pm 0.31
\label{eq:bsgTH}
\eeq 
where the theoretical uncertainty, $\sim10\%$, is less than half
the previous LO uncertainty.  It includes the uncertainty in the SM input parameters $m_t$, $\alphas (M_Z)$, $\alphae$, $m_c/\mb$, $\mb$, CKM angles and
the residual scale dependence uncertainties.  Nevertheless the error
is dominated by the uncertainty on the SM input parameters, as claimed
in Ref.~\cite{ref:bsgNLO2}.  Comparing with the experimental result in
Eq.~(\ref{eq:bsgEXP}), the SM prediction is higher than the observed
branching ratio, although the disagreement is less than two
experimental standard deviations.

To compute the deviations expected in the MGM, we look closer at the
theoretical calculation. The non-leptonic low energy effective
Hamiltonian $H^{\rm NL}_{\rm eff}$ relevant for the $\bsg$ decay is
\beq H^{\rm NL}_{\rm eff}=-\frac{4 G_F}{\sqrt{2}}V^{*}_{ts}V_{tb}
\sum_{i=1}^{8}C_i(\mub) Q_{i}(\mub)
\label{eq:heffNL}
\eeq where $V_{ij}$ are the CKM matrix elements, $C_i(\mub)$ the
Wilson coefficients and $Q_i(\mub)$ the corresponding operators
evaluated at the renormalization scale $\mub\simeq \O(\mb)$. Only the
magnetic and chromomagnetic dipole operators, $Q_7$ and $Q_8$
respectively, are significantly affected by new MGM amplitudes; their
expressions are given by 
\bea Q_7&=&\frac{e}{16 \pi^2}m_b\left(
\bar{s}_L\sigma^{\mu\nu} b_R\right) F_{\mu\nu}
\label{eq:q7}\\ 
Q_8&=&\frac{g_s}{16
\pi^2}m_b\left( \bar{s}_L\sigma^{\mu\nu} T^a b_R\right) G^a_{\mu\nu}
\label{eq:q8}
\eea 
where $T^a$ are the $\rm SU(3)_{color}$ generators and
$F_{\mu\nu}$ and $G_{\mu\nu}^a$ are respectively the electromagnetic
and $\rm SU(3)_{color}$ field strengths with $e$ and $g_s$ the
corresponding coupling constants.  The remaining operators in the
complete basis can be found in Ref.~\cite{ref:bsgNLO1}
or~\cite{ref:bsgNLO2}.  The Wilson coefficients $C_i(\mu)$, which
satisfy the RG equation \beq \mu \frac{d}{d \mu}
C_i(\mu)=C_j(\mu)\gamma_{ji}(\mu), \eeq can be expanded, together with
the anomalous dimension matrix $\gamma_{ij}$, in powers of $\alphas$
as follows: 
\bea
\hat{\gamma}&=&\frac{\alphas(\mu)}{4\pi}\hat{\gamma}^{(0)}+
\frac{\alphas^2(\mu)}{(4\pi)^2}\hat{\gamma}^{(1)}+ \O(\alphas^3)\\
C_i(\mu)&=&C_i^{(0)}(\mu)+\frac{\alphas(\mu)}{4\pi}C_i^{(1)}(\mu) +
\O(\alphas^2). \eea 
Then the solution in the $\ms$ renormalization
scheme for $\mu=\mub$ is given by 
\bea C_7^{(0)}(\mub) &=&
\eta^{\frac{16}{23}} C_7^{(0)}(\mw)+
\frac{8}{3}\left(\eta^{\frac{14}{23}}-\eta^{\frac{16}{23}}\right)C_8^{(0)}
(\mw)+ \sum_{i=1}^{8} h_i\eta^{a_i}
\label{eq:c7lo} \\ 
C_7^{(1)}(\mub) &=& \eta^{\frac{39}{23}} C_7^{(1)}(\mw )+
\frac{8}{3}\left(\eta^{\frac{37}{23}}-\eta^{\frac{39}{23}}\right)C_8^{(1)}
(\mw) \nl &+& \left(\frac{297664}{14283}
\eta^{\frac{16}{23}}-\frac{7164416}{357075} \eta^{\frac{14}{23}} +
\frac{256868}{14283} \eta^{\frac{37}{23}} - \frac{6698884}{357075}
\eta^{\frac{39}{23}} \right)C_8^{(0)}(\mw ) \nl &+&
\frac{37208}{4761}\left( \eta^{\frac{39}{23}}-\eta^{\frac{16}{23}}
\right)C_7^{(0)}(\mw ) +\sum_{i=1}^{8}\left( e_i \eta E(x_t)+f_i+g_i
\eta \right) \eta^{a_i}
\label{eq:c7nlo}
\eea where $\eta=\alphas (\mw )/\alphas (\mub)$, ~$x_t=m_t^2/\mw^2$,
and $\alphas$ is calculated at NLO. The SM contribution to
$C_{7,8}^{(0,1)}(\mw)$ together with the real numbers
$a_i,~f_i,~g_i,~h_i$, and the function $E(x)$ can be found in
Ref.~\cite{ref:bsgNLO1}.

Finally the branching ratio $\BRg$, conventionally normalized to the
semileptonic branching ratio $\BR(\BXcenu)=(10.4\pm0.4)\%$
~\cite{ref:pdg}, is given by~\cite{ref:bsgNLO1} 
\bea
\BRg^{\rm NLO}&=&\BR(\BXcenu ) \frac{|V_{ts}^{*} V_{tb}|^2}{|V_{cb}|^2}
\frac{6 \alphae}{\pi g(z) k(z)}\left(1-\frac{8}{3}\frac{\alphas
(\mb)}{\pi}\right) \left(|D|^2+A\right)(1+\delta_{np}) \nl D&=&
C_7^{(0)}(\mub)+\frac{\alphas (\mub)}{4\pi} \left(C_7^{(1)}(\mub)+
\sum_{i=1}^{8}C_i^{(0)}(\mub) \left[r_i(z)+\gamma_{i7}^{(0)}
\log{\frac{m_b}{\mub}}\right]\right) \nl A&=& \left( e^{-\alphas(\mub)
\log{\delta (7+2\log{\delta})/3\pi}}-1\right) |C_7^{(0)}(\mub )|^2 \nl
&+&\frac{\alphas (\mub)}{\pi}\sum_{i,j=1}^{8}C_i^{(0)}(\mub )
C_j^{(0)}(\mub ) f_{ij}(\delta)
\label{eq:BRg}
\eea 
where $z=m_c^2/m_b^2$ and in the last sum $i\le j$.
The expressions for all LO Wilson coefficients $C_i^{(0)}(\mub)$ together with the functions $g(z)$, $k(z)$, $r_i(z)$ and $f_{ij}(\delta)$ can be found in
Ref.~\cite{ref:bsgNLO1}.  The term $\delta_{np}$, of order a few
percent, includes the non-perturbative $1/\mb$ \cite{ref:bsgNPmb},
and $1/m_c$ \cite{ref:bsgNPmc},~\cite{ref:bsgllNPmc} corrections.

Some comments about this result are in order:
\begin{enumerate}
\item With NLO accuracy the inclusive $\bsg$ decay rate is given by
the sum of the total rate for $\bsg$ decay and its gluon
Bremsstrahlung corrections $\bsg g$, where in the former the
$\O(\alphas)$ radiative corrections to the matrix element are included
and in the latter an explicit lower cut on the photon energy
resolution $E_{\gamma}>(1-\delta)\mb/2$ is made.  In the sum of these
two contributions the only remaining infrared cutoff is the logarithm
of $\delta$ in the $A$ term of Eq.~(\ref{eq:BRg}).
\item The LO result can easily be recovered by taking the limit of
$\alphas \rightarrow 0$ inside the expression for the $D$ and $A$
terms.
\item To retain a strictly NLO result, the terms proportional to
$\alphas^2$ in the $|D|^2$ expression should be discarded.
\end{enumerate}

Since the new physics scale is above the electroweak scale, the new
contributions to the $\bsg$ decay will affect only the SM Wilson
coefficients at the $\mw$ scale. Therefore the non-perturbative
resummation of the large logarithms from $\mw$ to $\mub$ can be taken
completely from the existing SM calculation. Of course this would not
hold if there were new operators, generated at the electroweak scale,
which would significantly correct $C_7(\mub)$ via mixing with
$Q_7$. But in the MGM this does not occur: the dominant contributions
to the $\bsg$ decay enter only in the Wilson coefficients $C_7(\mw)$
and $C_8(\mw)$. We define fractional deviations $R_{7,8}$ from the SM
amplitudes: 
\beq C^{(0)}_{7,8}(\mw)\equiv C^{(0)
\rm SM}_{7,8}(\mw)\left(1+R_{7,8}\right)
\label{eq:param}
\eeq where $C^{(0) \rm SM}_i(\mw)$ represent the LO SM
contribution. Inserting these definitions into the $\BRg$ formula in
Eq.~(\ref{eq:BRg}) yields a general parametrization of the branching
ratio in terms of the new contributions~\cite{ref:short} : \bea 10^4
\BR^{\rm NLO}_\gamma &=& (3.48\pm 0.31) \left(1 + 0.622 R_7 + 0.090
R_7^2\right.\nonumber\\ &+& \left.0.066 R_8 + 0.019 R_7 R_8 + 0.002
R_8^2\right)
\label{eq:bsgPAR}
\eea 
where the following central values are used: 
$m_t^{\rm pole} \simeq m_t^{\rm \overline{MS}}(m_Z) \simeq 174\,\rm GeV$, 
$m_b^{\rm pole} = 4.8\,\rm GeV$, $m_c^{\rm pole} = 1.3\,\rm GeV$, 
$\mu_b = m_b$, $\alpha_s(m_Z) = 0.118$, $\alphae^{-1}(m_Z) = 128$, 
$\sin^2\theta_W = 0.23$ and a photon energy resolution corresponding 
to $\delta = 0.9$ is assumed.

What are the contributions to $R_{7,8}$ in the MGM model? In general
in the MSSM the diagrams which contribute to the $\bsg$ amplitude, in
addition to the SM, are given by a chargino, neutralino, gluino and
charged Higgs exchange.

The charged Higgs-exchange amplitude $(\cal{A}_{\rm H^{-}})$ is
well-known to be large~\cite{ref:bbmr},~\cite{ref:bsg1},~\cite{ref:bsg2}.  More
recently it was pointed out that for large $\tgb$ also the chargino and gluino
exchange can give sizeable contributions to the $\bsg$
amplitude~\cite{ref:bsg1}.  In particular, the leading large $\tgb$
contributions to the amplitude arise from exchanges of a pure higgsino
$(\cal{A}_{\rm \tilde{h}^{-}})$, a mixed wino-higgsino $(\cal{A}_{\rm
\tilde{W}\tilde{h}^{-}})$, or a gluino $(\cal{A}_{\rm \tilde{g}})$, in
descending order of importance.  Thus \beq R_{7} =\frac{\cal{A}_{\rm
H^{-}}+\cal{A}_{\rm \tilde{h}^{-}}+\cal{A}_{\rm
\tilde{W}\tilde{h}^{-}}+ \cal{A}_{\rm \tilde{g}}}{\cal{A}_{\rm SM}}
\label{eq:R7}
\eeq where the amplitude $\cal{A}_{\rm SM}$ correspond to an exchange
of a W boson.  We stress here that the same mechanism of $\tgb$
enhancement is present also in the $\delta m_b/m_b$ radiative
corrections to the bottom quark masses~\cite{ref:bsg1}.  Indeed the
diagrams contributing to the $\bsg$ amplitude are the same as for
$\delta m_b/m_b$, but where a photon line is attached in all possible
ways. Also, as explained above, the MSSM contributions are proportional to the relative sign of $\mu$, $M_{1/2}$ and $\tgb$. In the MGM this sign is calculable, and therefore a unique prediction of $\bsg$ decay rate in terms of only the messenger and gaugino mass can be given \cite{ref:rs},\cite{ref:short}.

In the MSSM the complete and exact analytical results for the $\bsg$
and $\bsglu$ total amplitudes can be found in Ref.~\cite{ref:bbmr}, in
terms of the various mass eigenvalues. However, since in the MGM the
squarks are quite heavy, a mass insertion approximation will suffice
\cite{ref:rs}. Here we diagonalize the chargino mass matrix exactly,
since we are interested in light charginos. The result for the partial
amplitudes is, up to an overall normalization 
%%%[of $\alpha_W\sqrt{\alpha}/(2\sqrt{\pi}\,m^2_W)$] 
~\cite{ref:short}:
\bea {\cal A}_{\rm SM} &=& \frac32 V_{ts} x_{tW} \left[\frac23
F_1(x_{tW}) + F_2(x_{tW})\right] \label{eq:ASM}\\
{\cal A}_{\rm H^-} &=& \frac12 V_{ts} r_b x_{tH} \left[\frac23
F_3(x_{tH}) + F_4(x_{tH})\right] \label{eq:A2H}\\
{\cal A}_{\rm \tilde h^-} &=& \frac12 V_{ts} r_b{\tan\beta} {\mu m_t
m_{\tilde t_L \tilde t_R}^2 \over m_{\tilde t_L}^2 m_{\tilde t_R}^2}
\left[M_2^2{F(x_{\tilde h_1 \tilde t},x_{\tilde h_2 \tilde t})\over
M_{\tilde h_1^-}^2 - M_{\tilde h_2^-}^2} - |m_{\tilde t_L} m_{\tilde
t_R}| {F'(x_{\tilde h_1 \tilde t},x_{\tilde h_2 \tilde t})\over
M_{\tilde h_1^-}^2 - M_{\tilde h_2^-}^2} \right]
\label{eq:AHH}\\
{\cal A}_{\rm \widetilde W \tilde h^-} &=& r_b{\tan\beta} {m_W^2
m_{\tilde t_L \tilde c_L}^2 \over m_{\tilde t_L}^2 m_{\tilde c_L}^2}
\mu M_2 {F(x_{\tilde h_1 \tilde q_L},x_{\tilde h_2 \tilde q_L}) \over
M_{\tilde h_1^-}^2 - M_{\tilde h_2^-}^2} \label{eq:AWH}\\
{\cal A}_{\rm \tilde g} &=& \frac89 r_b{\tan\beta}
{\alpha_s\over\alpha_W} {m_W^2 m_{\tilde t_L \tilde c_L}^2 \mu M_3
\over m_{\tilde q}^6} F_{\rm gl}(x_{M_3 \tilde q}) \,, \label{eq:AGL}
\eea
where $r_b \equiv 1/(1+{\delta m_b/m_b})$ accounts for the mass
corrections to $\mb$ \cite{ref:rs}.  The functions $F_i$ can be found
in Ref.~\cite{ref:bbmr}, while the new functions are defined as
follows: \bea F(x_1,x_2) &=& f(x_1) - f(x_2), ~~~F'(x_1,x_2) = x_1
f(x_1) - x_2 f(x_2),\\ f(x) &\equiv& \frac{d}{dx}(x F_3 + \frac23 x
F_4),~~~~ F_{\rm gl}(x) \equiv \frac12 \frac{d^2}{dx^2}(x^2 F_4).
\eea We also use: \bea x_{tW} &=& \frac{m_t^2}{m_W^2},~~~~x_{tH} =
\frac{m_t^2}{m_{H^-}^2},~~~ x_{\tilde h_i \tilde t} = \frac{M_{\tilde
h_i^-}^2}{|m_{\tilde t_L} m_{\tilde t_R}|},\\ x_{\tilde h_i \tilde
q_L} &=& \frac{M_{\tilde h_i^-}^2}{|m_{\tilde t_L} m_{\tilde c_L}|},
~~~ x_{M_3 \tilde q} = \frac{M_3^2}{m_{\tilde q}^2} \eea where
$M_{\tilde h_i^-}$ is the $i$th chargino mass eigenvalue and
$m_{\tilde q}$ is the average squark mass.  The left-right stop mass
insertion is $m_{\tilde t_L \tilde t_R}^2 = m_t A_t$ ($>0$), while the
left-left stop-scharm mass insertion is given by \cite{ref:rs} \beq
m_{\tilde t_L \tilde c_L}^2 \simeq +V_{ts} m_{\tilde q}^2
\frac{\lambda_t^2}{4\pi^2} \ln(\frac{M_M}{m_{\tilde t}}). \eeq

The contribution to $R_8$ can be obtained from Eq.~(\ref{eq:R7}) after
the following replacement in the amplitudes of
Eqs.~(\ref{eq:ASM}--\ref{eq:AGL}) : \beq \frac23 F_1 + F_2\to
F_1,\quad\frac23 F_3 + F_4 \to F_3, \quad f \to \frac{d}{dx}(x F_4)
\eeq \beq F_{\rm gl}(x)\to \frac{9}{16} \frac{d^2}{dx^2}(3 x^2 F_3
+\frac13 x^2 F_4).  \eeq

At present in the MSSM the supersymmetric corrections to $C_7(\mw)$ and
$C_8(\mw)$ are known only at LO; we will return to comment on the
resultant uncertainties when we present the numerical predictions for
$\BRg$ decay, in terms of the SUSY breaking scale and the messenger
mass scale $M_{M}$, in Sec. 6.
%%%%%%%%%%%%%%%%%%%%%%%%%%%%%%%%%%%%%%%%%%%%%%%%%%%%%%%%%%%%%%%%%%%%%%%%%%%

\section{ ${\bf \bsll }$ }
%%%%%%%%%%%%%%%%%%%%%%%%%%%%%%%%%%%%%%%%%%%%%%%%%%%%%%%%%%%%%%%%%%%%%%%%%%%
In this section we consider the semileptonic flavor-changing decays
$\bsee$, $\bsmumu$ and $\bstautau$. These processes are interesting
for several reasons: first, they are sensitive to new physics with
large $\tan\beta$ for the same reason that $\bsg$ is sensitive, and in
fact because of the same operator $Q_7$ (see also
Ref.~\cite{ref:hewells}); second, they involve other operators as
well, and so can serve as complementary tests of the
model~\cite{ref:agm}; and third, since there are several measurable
quantities (branching ratios and asymmetries in several systems), they
can potentially yield much more data.  At present these decays are
known at the NLO logarithmic accuracy for the
SM~\cite{ref:bsll1},~\cite{ref:bsll2}.  The $1/\mb$ non-perturbative
contributions are small and well under control, except near the
end-point of the dilepton mass spectrum~\cite{ref:ali}.  Recently it
was pointed out that other non-perturbative power corrections
$\O(\Lambda^2_{QCD}/m^2_c)$ could affect the amplitude, but in
practice they are very small~\cite{ref:bsgllNPmc}. Moreover there are
other non-perturbative effects due to the resonance regions in the
dilepton invariant mass distribution. But these do not have to be
taken into account if certain regions are excluded from the
integration region.

Unfortunately, the expected branching ratio for these decays is too
small to be observed in present experiments. The SM predictions for
the non-resonant decay rates $\BR(\bsll)\equiv \BRll$ are given
by~\cite{ref:bsllsusy},~\cite{ref:short} \bea \BRee&\simeq&
7.0\times 10^{-6}
\label{eq:bseeSM} \\
\BRmm&\simeq& 4.5 \times 10^{-6}
\label{eq:bsmumuSM} \\
\BRtt&\simeq& 2.0 \times 10^{-7};
\label{eq:bstautauSM}
\eea at present, experiments only place some upper bounds on $\BRee$
and $\BRmm$, $\O(\mbox{few}\times 10^{-5})$, or about an order
of magnitude above the SM~\cite{ref:bsllexp}.  However an increase of
two to four orders of magnitude in the experimental sensitivity is
expected in the near future at the Tevatron~\cite{ref:bsllfut}.  With
such statistics it will be possible to measure the branching ratio and
the energy or angular distributions, even for the more rare $\tau$
decay mode.

We now turn to our prediction for the branching ratios and
asymmetries. The effective Hamiltonian $H_{\rm eff}$ relevant for the
$\bsll$ decay is: \beq H_{\rm eff}=H_{\rm eff}^{\rm
NL}-\frac{G_F}{\sqrt{2}}V^{*}_{ts}V_{tb} \left\{ C_9(\mub)Q_9(\mub)+
C_{10}(\mub)Q_{10}(\mub)\right\}
\label{eq:heffSL}
\eeq where $H_{\rm eff}^{\rm NL}$ is given in Eq.~(\ref{eq:heffNL}) with
$\mub\simeq \O(m_b)$, the magnetic-dipole operator $Q_7$ is defined in
Eq.~(\ref{eq:q7}) and the semileptonic operators $Q_9$ and $Q_{10}$ are 
given by
\bea
Q_9&=&\left(\bar{s}\gamma^{\mu}(1-\gamma_5)b\right)\left(\bar{l}\gamma^{\mu}
l\right)\label{eq:q9}\\
Q_{10}&=&\left(\bar{s}\gamma^{\mu}(1-\gamma_5)b\right)\left(\bar{l}\gamma^{\mu}
\gamma_5 l\right).  
\label{eq:q10}
\eea 
The Wilson coefficients $C_9(\mub)$ and
$C_{10}(\mub)$ at NLO are: \beq
C_9(\mub)=\frac{\alpha}{2\pi}\left(P_0(\mub,\eta)+P_E E\right)
+C_9(\mw) \eeq where the initial conditions $C_{9,10}(\mw)$ in the HV
scheme are given by~\cite{ref:bsll2} 
\bea
C_9(\mw)&=&\frac{\alpha}{2\pi}\left(\frac{1}{\sww}Y -4Z\right)
\label{eq:c9}\\
C_{10}(\mub)&=&C_{10}(\mw)=-\frac{\alpha}{2\pi\sww}Y.  
\label{eq:c10}
\eea
The terms $Y$, $Z$ and $E$ contain the results of the one-loop
integration for the $\gamma,Z$ penguin and box diagrams at the
electroweak scale. The SM expressions for $Y,Z,E$, respectively the
functions $Y^{\rm SM}(x_t)$, $Z^{\rm SM}(x_t)$ and $E^{\rm SM}(x_t)$
($x_t=m_t^2/\mw^2$), can be found in
Refs.~\cite{ref:bsll1},~\cite{ref:bsll2}.  The function $P_0$ includes
the resummation of the large logarithms at the NLO and it is scheme
dependent.  Finally, in the limit $\eta\rightarrow 1$, $P_E\rightarrow
0$ and $P_0 \rightarrow 0$, so that the initial condition
$C_9(\mu)\rightarrow C_9(\mw)$ is recovered.  The expressions for
$P_0$ and $P_E$, in both HV and NDR scheme for the former, can be
found in Ref.~\cite{ref:bsll2}.

Note that $C_{10}(\mub)$ does not depend on the renormalization scale
$\mub$ [see Eq.~(\ref{eq:q10})]. 
This is a consequence of the P and C symmetry of the
electromagnetic current which forbids any mixing of $Q_{10}$ with
$Q_9$ or any other operator of the non-leptonic effective Hamiltonian
$H^{\rm NL}_{\rm eff}$.

At NLO accuracy the total amplitude for $\bsll$ decay can be expressed
as a function of $C_9^{\rm eff}(\mub)= C_9(\mub)+C_9^{ME}(\mub)$,
$C_7^{(0)}(\mub)$ and $C_{10}(\mw)$ where $C_9^{ME}(\mub)$ contains
the $\O(\alphas)$ one-loop matrix elements needed to cancel the
leading $\mub$ dependence in $C_9(\mub)$. Moreover $C_9^{\rm eff}$ is
scheme independent since $C_9^{ME}$ cancels exactly the scheme
dependence of $C_9(\mub)$ ~\cite{ref:bsll2}.  Note that at NLO
accuracy, $C_7(\mub)$ enter only at the leading order.

Finally the differential decay rate for $b\to s\ell^+\ell^-$ is given
by
\bea {d^2 \Gamma_{\ell\ell}\over dy_+ dy_-} &=& \frac{G_F^2}{256\pi^5}
m_b^5 |V^{*}_{ts} V_{tb}|^2 \alphae^2 \,\left[\left|C_9^{\rm
eff}(\mub)\right|^2 K^{9,9} + C_{10}^2(\mw) K^{10,10}
\right.\nonumber\\ &+&\left. \left(C_7^{(0)}(\mub)\right)^2 K^{7,7} +
C_7^{(0)}(\mub) {\rm Re}\left[C_{9}^{\rm eff}(\mub)\right] K^{7,9}
\right.\nonumber\\ &+& \left. C_{10}(\mw) \left( C_7^{(0)}(\mub)
K^{7,10} + {\rm Re}\left[C_{9}^{\rm eff}(\mub)\right]
K^{9,10}\right)\right].
\label{eq:dGam}
\eea
Note that $C^{(0)}_7(\mub)$ is real [see Eq.~(\ref{eq:c7lo})] as is
$C_{10}(\mw)$, while $C_9^{\rm eff}(\mub)$ is complex --- its
imaginary part originates in the one-loop matrix elements in
$C_9^{ME}(\mub)$. The kinematic variables are $y_{\pm} = 2
E_{\pm}/m_b$ where $E_{\pm}$ are the lepton energies measured in the
$b$ rest frame. We prefer to use the following combinations: \bea \hat
s &=& y_+ + y_- - 1\\ \nonumber \hat y &=& y_+ - y_- = -\hat y_{\rm
max} \cos\theta \eea where $\hat{y}_{\rm max} = (1 - \hat s) \sqrt{1 -
4 x_\ell/\hat s}$, $x_\ell = m_\ell^2/m_b^2$ and $\theta$ is the angle
between the $b$ and the $\ell^+$ in the $\ell^+\ell^-$ rest frame.
Including only the lepton mass corrections, we find
\bea K^{9,9} &=& \frac12 \left[1 - \hat s^2 - \hat y^2 + 4 x_\ell (1
-\hat s) \right] \label{eq:Knini}\\ K^{10,10} &=& \frac12 \left[1
-\hat s^2 - \hat y^2 - 4 x_\ell (1 - \hat s) \right]
\label{eq:Ktete}\\
K^{7,7} &=& \frac{2}{\hat s} \left[1 - \hat s^2 + \hat y^2 +
{4x_\ell\over\hat s} (1 - \hat s) \right]
\label{eq:Ksese}\\ 
K^{7,9} &=& 4 \left[1 - \hat s + {2 x_\ell\over\hat s} (1 - \hat s)
\right]
\label{eq:Kseni}\\ 
K^{7,10} &=& 4 \hat y
\label{eq:Ksete}\\ 
K^{9,10} &=& 2 \hat s \hat y .
\label{eq:Knite}
\eea The results of
Eqs.~(\ref{eq:dGam}),~(\ref{eq:Knini}--\ref{eq:Knite}) are in
agreement with Ref.~\cite{ref:bsllsusy}, where the massless case is
considered.  Moreover, after integrating Eq.~(\ref{eq:dGam}) with respect
to $\hat{y}$, our result agrees with that of Ref.~\cite{ref:hewett},
including the mass corrections.

The $1/\hat{s}$ terms present in Eqs.~(\ref{eq:Ksese},\ref{eq:Kseni})
are due to the photon propagator which connects the quark matrix
element of the magnetic dipole operator $Q_7$ to the electromagnetic
lepton current.  As observed in Ref.~\cite{ref:bsllsusy} the last term
of Eq.~(\ref{eq:Ksese}) proportional to the lepton mass ($x_\ell$)
can not be completely neglected even in the light lepton case. In
particular it gives a finite contribution to the total branching ratio
(roughly a few per cent) even in the $m_l^2\rightarrow 0$ limit.
Indeed the $1/\hat{s}^2$ term in Eq.~(\ref{eq:Ksese}) generates a
pole $\O(1/m^2_l)$ in the integrated branching ratio which is
cancelled by the $\O(m^2_l)$ terms present in the numerator.

The total branching ratio for a particular $\ell$ is obtained
integrating $d^2 \Gamma_{\ell\ell}/dy_+ dy_-$ over the entire
kinematically-allowed range for that $\ell$.   
The forward-backward asymmetry $\All$ (which is the same as the energy asymmetry) is proportional to the integral over the range $y_- > y_+$ (i.e.,
$\cos\theta > 0$) minus the integral over the range $y_- < y_+$
($\cos\theta < 0$). The allowed kinematic limits are $-\hat y_{\rm
max} < \hat y < +\hat y_{\rm max}$ and $4 x_\ell < \hat s < 1$.

To avoid intermediate charmonium resonances and non-perturbative
phenomena near the end point, we exclude the same ranges of $\hat s$
specified in Ref.~\cite{ref:bsllsusy}.  In particular, the integrating
regions of $\hat{s}$ in the $l=e, \mu$ case are \bea \left(\mb^2
\hat{s}\right) &\in& \left\{4 m_{l}^2,~(2.9~\mbox{GeV})^2 \right\}
\cup \left\{(3.3~\mbox{GeV})^2,~(3.6~\mbox{GeV})^2 \right\} \cup \nl
&& \left\{(3.8~\mbox{GeV})^2,~(4.6~\mbox{GeV})^2 \right\}
\label{eq:rangeemu}
\eea 
and for $l=\tau$ \beq \left(\mb^2\hat{s}\right) \in \left\{4
m_{\tau}^2,~(3.6~\mbox{GeV})^2\right\} \cup
\left\{(3.8~\mbox{GeV})^2,~(4.6~\mbox{GeV})^2 \right\}
\label{eq:rangetau}
\eeq

\begin{figure}
\centerline{\epsfig{figure=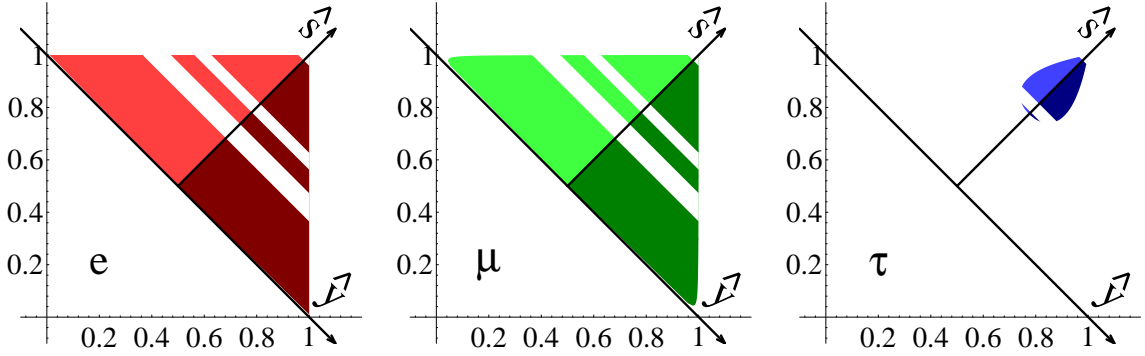,height=5cm,angle=0}}
\caption{
{\small
Dalitz plots for the integrating regions of $\bsll$ decays corresponding
to $e,~\mu$ and $\tau$ decay modes from the left to the right respectively.}}
\label{fig:dalplot}  
\end{figure}

The Dalitz plots for the decays $\bsll$ are shown in Fig. \ref{fig:dalplot}. The dimensionless lepton energies $y^+$ and $y^-$ run along the horizontal and vertical axes, respectively. The rotated axes correspond to $\hat y$ (pointing right and down) and $\hat s$ (pointing right and up). Only the regions of $\hat s$ included in the integration [see Eqs.~(\ref{eq:rangeemu},\ref{eq:rangetau})] have been shaded. The branching ratios are obtained by adding the integrals over the (colored) dark- and light-shaded areas, while the asymmetries require subtracting the integral over the light-shaded region from the integral over the dark-shaded region.

As usual, we normalize the branching ratio to the semileptonic decay
rate $b\to c\,e\overline\nu_e$ in order to eliminate the large
dependence on the $b$ quark mass: $\Gamma_{\ell\ell} \to
\widehat\Gamma_{\ell\ell} \equiv \Gamma_{\ell\ell}/\Gamma_{b\to c}$,
hence
\bea {\rm BR}_{\ell\ell} &=& ({\rm BR}_{b\to c}) \int d\hat s
\,2\!\!\int_0^{\hat y_{\rm max}} d\hat y \left({d^2
\widehat\Gamma_{\ell\ell}\over d\hat s d\hat y}\right)_{\rm symm}
\label{eq:BRll} \\
\All &\equiv& {N(y_- > y_+) - N(y_+ > y_-) \over N(y_- > y_+) + N(y_+ > y_-)}
\nonumber\\ &=& -{1\over\widehat\Gamma_{\ell\ell}} \int d\hat s
\,2\!\!\int_0^{\hat y_{\rm max}} d\hat y \left({d^2
\widehat\Gamma_{\ell\ell}\over d\hat s d\hat y}\right)_{\rm antisymm}
\label{eq:All}
\eea
in which the symmetric (antisymmetric) rate must include only the
kinematic coefficients symmetric (antisymmetric) in the exchange 
$\hat y\rightarrow -\hat y$.

We have also calculated some polarization asymmetries, namely the rate when 
one of the lepton's spin is parallel to its momentum versus the rate when that 
spin is antiparallel to the momentum. We have considered both the integral of 
the asymmetry in the differential rates, and the asymmetry in the integrated 
rates, and also looked at the three possible polarization directions, 
using the work of ref.~\cite{ref:KS}.

Let now discuss the impact of the MGM model in the $\bsll$ decays.

The contribution of the MSSM to $C_{9,10}(\mw)$ is given by the
$\gamma$ and $Z$ penguin diagrams proportional to the form factor
$\gamma_{\mu}$, and by the box diagrams. Their contribution are
incorporated into the $Y$, $Z$ and $E$ expressions. Inside the penguin
and box diagrams can run the charged Higgs, charginos, gluinos and
neutralinos. Since $P_E$ is two orders of magnitude smaller than
$P_0$, and the SUSY contribution to $E$ should be smaller or
comparable to the SM one, we will neglect the $E$ contribution in our
analysis. The complete analytical results for $Y^{\rm SUSY}$ and
$Z^{\rm SUSY}$ can be found in
Refs.~\cite{ref:bbmr},~\cite{ref:bsllsusy}.

In the MGM model we have computed the SUSY contribution due to
chargino, gluino and neutralino exchange in the $\gamma$ and $Z$
penguin diagrams and box diagrams and we find that $Y^{\rm SUSY}$ and
$Z^{\rm SUSY}$ are very small comparing to the leading SM
contributions, roughly a few percent. (Note that the squarks are quite
heavy, but there is no compensating large $\tan\beta$
enhancement.) 
Also the charged-Higgs contribution to $Y$ and $Z$ is small, 
being suppressed by at least ${\cal O}(m_s/m_b)$ and in practice it is 
of order of a few percent in both the penguin and box diagrams.

New chirality-violating four-fermion operators could be generated at
the squark mass with their chiral suppression partially compensated for by
$\tgb$. However we found that their contribution is small compared to
the leading SM contribution to $C_{9,10}(\mw)$, at least for large
squark masses \O(0.5--1 TeV).

Thus we find that the only relevant effect of the MGM on the $\bsll$
decays can be parametrized by $R_7$ in Eq.~(\ref{eq:R7}). In particular
for the ${\rm BR}_{ll}$ and $\All$ we find~\cite{ref:short}
\begin{eqnarray}
10^7\,{\rm BR}_{ee} &=& 68.5 + 22.4 R_7 + 6.1 R_7^2 \label{eq:BRee} \\
{\rm A}_{ee} &=& (4.52 - 3.01R_7) / (10^7 \,{\rm BR}_{ee})
\label{eq:Aee} \\ 10^7\,{\rm BR}_{\mu\mu} &=& 44.64 + 2.46 R_7 + 1.81
R_7^2 \label{eq:BRmumu} \\ {\rm A}_{\mu\mu} &=& (4.68 - 2.90 R_7) /
(10^7 \,{\rm BR}_{\mu\mu}) \label{eq:Amumu} \\ 10^7\,{\rm
BR}_{\tau\tau} &=& 2.013 - 0.201 R_7 + 0.009 R_7^2 \label{eq:BRtautau}
\\ {\rm A}_{\tau\tau} &=& (0.434 - 0.042 R_7) / (10^7 \,{\rm
BR}_{\tau\tau}) \label{eq:Atautau} \,.
\end{eqnarray}

The coefficients of these $R_7$ polynomials were obtained using the
same central values for the SM parameters used in the $\bsg$ decay.
We have checked the sensitivity of these predictions to the various
input parameters and to the scale $\mub$.  The main sources of
uncertainty are $m_t$, $\alpha_s(m_Z)$ and the residual $\mub$
dependence. We find that these uncertainties affect the normalizations
of the ${\rm BR}_{\ell\ell}$ by up to $\O(10\%)$. However the
uncertainties on the functional dependence on $R_7$ are much less, at
the percent level.  There is a small sensitivity to $R_8$ (which
nonetheless can be included). The normalizations of ${\rm A}_{ee}$ and
${\rm A}_{\mu\mu}$ are even more sensitive than the branching ratios
to $\mub$, but an improved SM NLO calculation can reduce the
uncertainties to less than $\sim10\%$. On the contrary the $\tau$
forward-backward asymmetry $\Att$ has a weaker functional
dependence on $R_7$ then the other decay modes.  
Moreover we find that the various integrated polarization asymmetries 
are not particularly sensitive -- at most at the few percent level -- 
to $R_7$, 
so they would not serve to test this model and we do not include them in our 
final results.

Finally we would like to stress that, when the present residual $\mub$
dependence is reduced by improving the SM NLO calculation, and when
more precise measurements of $m_t$ and $\alpha_s(m_Z)$ are available,
then the MGM model predictions for $\BRll$ and ${\rm A}_{ll}$ can be
significantly sharpened.

%%%%%%%%%%%%%%%%%%%%%%%%%%%%%%%%%%%%%%%%%%%%%%%%%%%%%%%%%%%%%%%%%%%%%%%%%%%%
%\newpage
\section{${\bf g_{\mu}-2}$}

The anomalous magnetic moment of the muon $\amu \equiv (g_{\mu}-2)/2$
is one of the most important high precision experiments, providing
extremely precise tests of QED and electroweak interactions as well as
strong contraints on new physics models. The theoretical prediction is
known to $\O(\alpha^5)$ in QED~\cite{ref:gmtwoGR}, and recently also
the two-loop electroweak radiative corrections have been
included~\cite{ref:gmtwoEW}.  The agreement between the theoretical
prediction and the experimental result is impressive -- but not exact.
The current average experimental value of $\amu^{\rm exp}$ is
~\cite{ref:gmtwoEXP} \beq 10^{10}\amu^{\rm exp} - (11~659~000) = 230
\pm 84
\label{eq:g2EXP}
\eeq The E821 experiment at Brookhaven National Laboratory (BNL) is expected to measure $\amu$ to within $\pm 4\times 10^{-10}$, or perhaps even $\pm 1-2\times 10^{-10}$ \cite{ref:gmtwoFUT}. The most up to date theoretical prediction in the SM is given by the sum of the following contributions \cite{ref:gmtwoSM}: \beq 
10^{10} \amu^{\rm SM}=\amu^{\rm QED}+\amu^{\rm EW}+\amu^{H(1)}(\mbox{vac
pol})+\amu^{H(2)}(\mbox{vac pol})+ \amu^{H}(\gamma\times \gamma)
\label{eq:g2THA}
\eeq 
where 
\bea 
\amu^{\rm QED}&=& 11~658~470.57~(0.19) \label{eq:g2THqed} \\ 
\amu^{\rm EW}&=& 15.1~(0.4) \label{eq:g2THew} \\ 
\amu^{H(1)}(\mbox{vac pol})&=& 701.1~(9.4) \label{eq:g2THh1} \\ \amu^{H(2)}(\mbox{vac pol})&=& -10.1~( 0.6) \label{eq:g2THh2} \\ \amu^{H}(\gamma\gamma)&=& -7.92~( 1.54) \label{eq:g2THgg}\,.
\eea 
$\amu^{\rm QED}$ includes \cite{ref:gmtwoQED} the most recent pure QED contributions at order $\alpha^5$; $\amu^{\rm EW}$ includes the
two-loop \cite{ref:gmtwoEW} electroweak corrections; the
$\amu^{H(1,2)}(\mbox{vac pol})$ include respectively the one-loop
\cite{ref:gmtwoHAD1} and two-loop \cite{ref:gmtwoHAD2} hadron vacuum
polarization corrections extracted with dispersion
relations from the measurement of the $e^+e^- \rightarrow
\mbox{hadrons}$ cross section; and $\amu^{H}(\gamma \gamma)$ contains the
hadronic ``light-by-light scattering'' contribution \cite{ref:gmtwoGG}.

The hadronic contribution to the vacuum polarization is the largest
source of error, reflecting the large experimental uncertainty of the
data.  However with the new experiments of BEPC at Beijing, DA$\Phi$NE
at Frascati and VEPP-2M at Novosibirsk the theoretical error in
$\amu^{H}(\mbox{vac pol})$ will be significantly reduced \cite{ref:gmtwoexpH}.  The hadronic light-by-light contribution arises from a different class of diagrams which cannot be directly extracted from experimental data, and
therefore a somewhat model-dependent calculation is used.

Finally, by combining the results of Eqs.~(\ref{eq:g2THA}--\ref{eq:g2THgg}) and summing the errors in quadrature we arrive at the following SM prediction:
\beq 
10^{10} \amu^{\rm SM}-(11~659~000)~=~170 \pm 10\,.
\label{eq:g2THN}
\eeq
The improved experimental precision in the next experiment at BNL and
the reduction of the theoretical uncertainties in the hadronic vacuum
polarization will allow a direct test of the electroweak corrections.

The expected deviations from $\amu^{\rm SM}$ induced by the MSSM, and the degree to which the MSSM can be constrained by improved experimental measurements, has been studied since the early days of supersymmetric phenomenology \cite{ref:gmtwoSUSYgen}. More recently the impact of large $\tan\beta$ was considered, first in the context of gravity-mediated SUSY breaking \cite{ref:gmtwoSUSY}, next in the general MSSM \cite{ref:moroi}, and finally specifically in gauge-mediated scenarios \cite{ref:cgw}. However, in this last work, the crucial parameters $\mu$, $\tgb$ and the sign of the new amplitudes were not predicted by the model. In contrast, in the MGM model we consider all three of these crucial parameters are predicted along with the rest of the MSSM spectrum in terms of our two fundamental parameters, allowing us to correlate our $\amu$ prediction with the predictions of $\bsg$ and $\bsll$.

The MSSM contributes to $\amu$ mainly via magnetic-dipole
penguin diagrams, with an exchange of a chargino or a neutralino
in the loop. We have calculated these contributions and our analytical results are in agreement, in magnitude and sign, with the results of
Ref.~\cite{ref:moroi} (after accounting for the erratum).

In the MGM model the chargino amplitudes dominate the neutralino ones,
and within the chargino contribution the terms proportional to the
chargino mass completely dominate.  Using also $v_D\ll v_U$ yields \cite{ref:short}, for the deviation from the standard model,
\begin{equation}
\Delta a_\mu^{\rm MGM} \simeq a_\mu^{\tilde h^-} \simeq
{3\alpha_2\over4\pi}\tan\beta {m_\mu^2 \mu M_2 F_\mu(x_{\tilde h_1
\tilde\nu},x_{\tilde h_2 \tilde\nu}) \over m_{\tilde \nu}^2 (M_{\tilde
h_1^-}^2 - M_{\tilde h_2^-}^2)}
\label{eq:Agmu}
\end{equation}
where 
\bea F_\mu(x_1,x_2) &=& f_\mu(x_1) - f_\mu(x_2),\\ \nonumber
f_\mu(x) &=& \frac{3 - 4 x + x^2 + 2 \ln x}{3 (1 - x)^3}, 
\eea 
and $x_{\tilde h_i \tilde\nu} = M_{\tilde h_i^-}^2/m_{\tilde \nu}^2$.
For our numerical results we will use the complete expressions for $a_\mu^{\rm MGM}$.

\section{MGM model predictions}

As explained in Sec. 2, when the self-consistent equation Eq. (\ref{eq:tgb}) for $\tgb$ is numerically solved, all of the relevant SUSY spectrum and physical quantities can be predicted as a function of only two fundamental parameters: the effective SUSY-breaking scale $\Lambda$ or equivalently the weak gaugino mass $M_2$, and the logarithm of the common messenger mass $M_M$. In principle these results will depend on the doublet and triplet messenger mass splitting, but the dependence is very weak for the expected small splitting. In all our computations we will assume that $M_{M_3}=1.3 M_{M_2}$.

We consider the range $2 \Lambda < M_M < 10^4\Lambda$. Messenger masses much nearer to the absolute lower bound $\Lambda$ (below which their scalar components would develop VEVs) are not of much interest: they are either already ruled out by vacuum stability constraints \cite{ref:rs} or require extremely heavy superpartners and thus mimic the standard model. Messenger masses much above ${\cal O}(10^4\Lambda)$ may also destabilize the vacuum or run into cosmological difficulties with a heavy gravitino (\cite{ref:rs,ref:gravbounds} and references therein), though this is by no means an airtight bound.

Within this range of $M_M$, the one-step solution we have used for the RG equations is an adequate approximation. In the low $M_M$ region, $\tgb$ is $\sim 50$ and there are strong cancellations in the $B$ term and in the $\bsg$ amplitudes, so an accurate solution is required. But as shown in Ref.~\cite{ref:rs}, the one-step approximation in this case is as accurate as keeping only the leading threshold corrections, and both of these approximations are entirely sufficient for our purposes.  For the highest values of $M_M$ we consider, the one-step approximation is not as accurate, but the cancellations in $B$ and in $\bsg$ are not as severe, so the reduced accuracy is once again sufficient. In the end, we estimate that the uncertainties in our predictions of $\BRg$ and $a_{\mu}$ are comparable to the corresponding SM ones.

\begin{figure}[t]
\centerline{\epsfig{figure=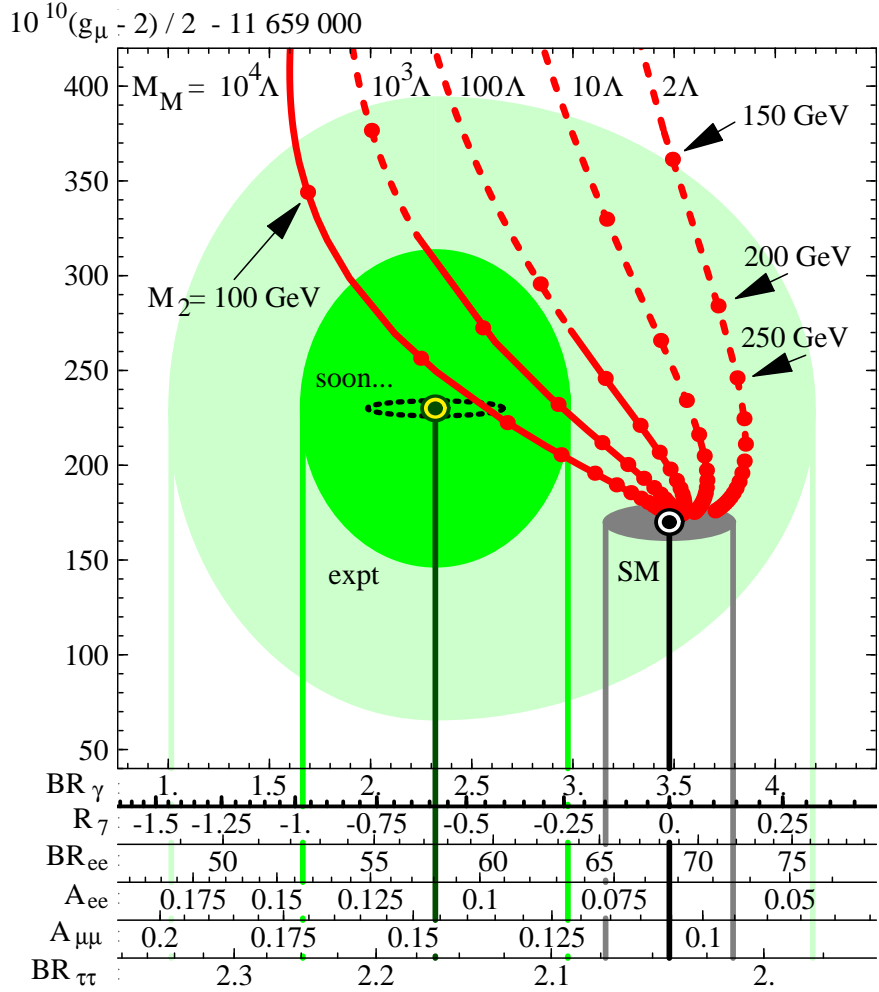,height=15cm,angle=0}}
\caption{
{\small
The $N=1$ MGM predictions (as heavy red curves) of $\BRg$ ($\times 10^4$) and 
$10^{10} (g_\mu-2) - 11 659 00$ as functions of the wino mass $M_2$, 
for various messenger masses $M_M$ and with heavy red dots at 
discrete $M_2$ values. The experimental and SM central values, 
indicated with circled dots, are surrounded by the corresponding 
uncertainty ellipses.
The extra horizontal axes are the correlated predictions for 
$R_7$, $\BRll$ ($\times 10^7$) and $\All$. }}
\label{fig:main1}  
\end{figure}

\begin{figure}[t]
\centerline{
\epsfig{figure=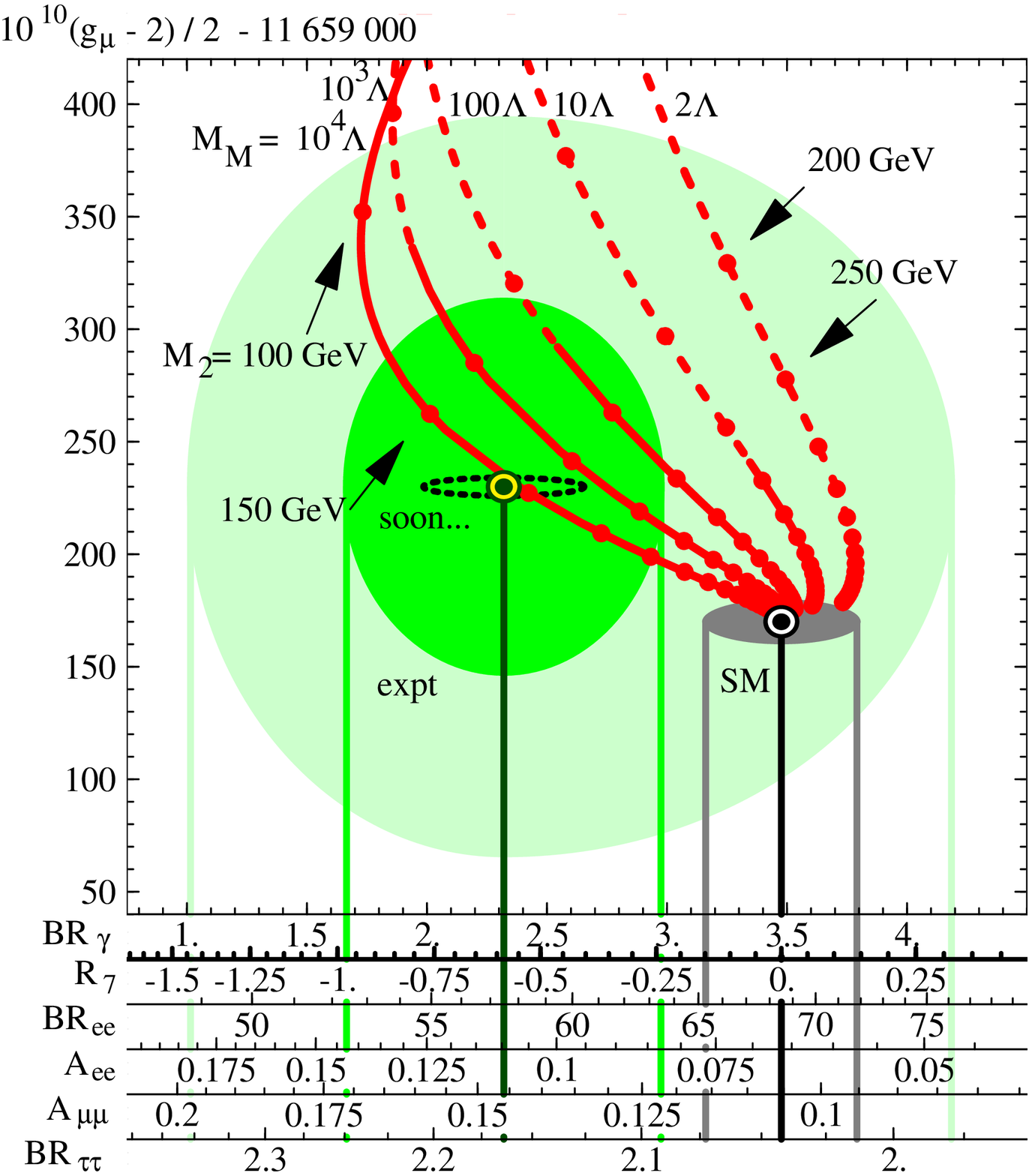,height=15cm,angle=0}}
\caption{{\small The MGM predictions as in Fig.~\ref{fig:main1}
but for $N=2$.}}
\label{fig:main2}  
\end{figure}

In Figs. \ref{fig:main1} and \ref{fig:main2} we present our MGM model predictions as a function of $M_M$ and the weak gaugino mass $M_2$ for $N=1$ or 2 messenger families, respectively. Different curves are given for different messenger masses $M_M$, and heavy (red) dots indicate discrete values of $M_2$. The SM-predicted and the measured central values are shown as circled dots. The dark and light green areas surrounding the experimental central value indicate the regions allowed at $1\,\sigma$ and $95 \%$ CL, respectively. The dark ellipse surrounding the central value of the SM predictions indicates the expected theoretical uncertainty, which should also be applied to each point on the MGM prediction curves. Finally, the small dashed ellipse centered (for the present) on the experimental central value indicates the $1\,\sigma$ uncertainties expected in the near future from BNL and CLEO.

The heavy (continuous and dashed) red curves are our predictions for 
$\amu$ along the vertical axis, and either $\BRg$ or any of the various $B\rightarrow X_s l^+l^-$ branching ratios and asymmetries. All the latter are correlated in their dependence on the single parameter $R_7$ which measures the deviation from the SM of the amplitude of the $Q_7$ operator. At present $\BRg$ is by far the best measure of $R_7$, which is why we emphasize it in our graphs. (The various quantities along the horizontal axes depend slightly and differently on the $R_8$ parameter, so to present them in the same graph we have chosen a specific value of $R_8 = 0$. But we estimate that this approximation affects the $R_7$ predictions by an uncertainty that is well within the SM one shown in the figure.) The curves become dashed for those ranges of $M_M$ and $M_2$ which are in fact ruled out by vacuum stability constraints \cite{ref:rs}, primarily due to the stau developing a charge-breaking VEV.

Several features of the MGM model predictions stand out from these figures: 
\begin{enumerate}
\item When the messengers are relatively heavy, the MGM predictions are in better agreement with experiment than those of the SM (except for extremely light superpartners): when the superpartners are lowered (to their natural values, around the electroweak scale) and the MGM departs from the SM, $a_\mu$ increases and $\BRg$ decreases towards the experimental central value. In particular, the sign of this effect is a prediction of the MGM.  
\item When the messengers are quite light, a fortuitous cancellation between the chargino and charged-Higgs amplitudes in $R_7$ prevents $\BRg$ and the various $B\rightarrow X_s l^+l^-$ branching ratios and asymmetries from deviating significantly from the SM, although $\amu$ is still raised as before.
\item The expected reduction in experimental uncertainty will tremendously sharpen the comparison with the SM and MGM predictions, possibly deciding between the two and definitely strongly constraining the latter.
\item At present, the (absolute) bounds vaccum instability bounds on $M_M$ and $M_2$ are stronger then the (95\% CL) experimental bounds due to $\BRg$ and $\amu$.
\end{enumerate}

As a function of $N$ the sfermion masses scale like $1/\sqrt{N}$ relative to the gauginos [see Eqs.~(\ref{eq:squarksleptonmass1})--(\ref{eq:squarksleptonmass2})].
Raising $N$ for fixed gaugino masses thus lowers the sfermions and amplifies the MGM effects. While this is the general trend, in practice there are some complications because of cancellations and the variation in $\tgb$. By comparing Figs. \ref{fig:main1},~\ref{fig:main2} throughout the allowed region
for fixed values of $M_M$ and $M_2$, we see that when $N$ is increased, $R_7$ (or equivalently  $\BRg$) is decreased and $a_{\mu}$ is increased. Raising $N$ is roughly equivalent to lowering $M_2$ while raising $M_M$. 

From the horizontal axes below those of $\BRg$ and $R_7$ in Figs. \ref{fig:main1},~\ref{fig:main2}, our predictions for $\BRee$, $\BRtt$, $\Aee$ and $\Amm$ may be read off. We do not give the results for $\BRmm$ and $\Att$ 
because they are not very sensitive to $R_7$. Evidently the MGM model can produce large deviations from the SM in $\Aee$ and $\Amm$, and somewhat smaller ones in $\BRee$ and $\BRmm$. Note however that the $\bsll$ quantities have their own theoretical uncertainties beyond those indicated in the figures,
mainly from the residual $\mub$ dependence, as explained in Sec. 4.
These could be reduced by improving the SM NLO (in $\alphas$) calculations,
at which point the MGM model would predict the branching ratios and asymmetries more precisely. We expect that at least some of these predictions could be tested at hadronic colliders in the next few years.

%\newpage
\section{Conclusions}

Within the minimal gauge-mediated SUSY-breaking model, which can naturally generate a large $\tgb$, we have computed the inclusive branching ratio $\BXsg$, the inclusive branching ratios and asymmetries for $\BXsll$ 
(with $\ell=e,~\mu,~\tau)$, and the  anomalous magnetic moment of the muon $a_{\mu}\equiv (g_{\mu}-2)/2$. In particular we included the complete next-to-leading order accuracy in the strong coupling for the B decays.

The MGM model, described in Sec. 2, has a naturally high $\tgb$ signature
provided that the tree-level soft-breaking $B$ parameter, which couples
the Higgs doublets in the scalar potential, vanishes.
Moreover this model is highly predictive: all the SUSY soft-breaking 
terms, as well as $\tgb$ and the physical relative sign (often called the sign of the $\mu$ parameter), can be predicted in terms of only two fundamental parameters: the SUSY breaking scale (or equivalently the wino mass), and the logarithm of a common messenger mass. Therefore the MGM predictions for these processes are strongly correlated.

The correlated predictions are shown in Figs. \ref{fig:main1} and \ref{fig:main2}. Note that the correlation between $\amu$ and $\BRg$ depends on the model parameters, $M_M$ or $M_2$, whereas the various b-quark-related quantities are all completely correlated through their dependence on the single quantity $R_7$ independent of $M_2$ and $M_M$ (a quite general feature of the MSSM, as discussed in Ref.~\cite{ref:hewells}). 
We have gone slightly beyond the minimal model by including the results 
for $N=2$ messenger families. We included the results of the numerical 
predictions for $N=1$ and $N=2$ messenger families in Figs.~\ref{fig:main1}
and~\ref{fig:main2} respectively. 
The general trend of the MGM predictions emerging from these results can be summarized as follows:
\begin{itemize}
\item When the messenger scale is a few orders of magnitude above $\Lambda \sim {\cal O}(100\,\rm TeV)$, and the superpartners are near their natural electroweak scale, the MGM predicts a higher anomalous magnetic moment and lower $b\to s\gamma$ branching ratio than the SM, in agreement with present measurements.
\item When the messengers are not far above $\Lambda$, the superpartners must be significantly higher than the electroweak scale, and (due to a cancellation) the $b\to s\gamma$ rate is largely unperturbed from its SM value, though the anomalous magnetic moment is raised.
\item Increasing the number of messenger families amplifies these predictions somewhat: it is similar to lowering the superpartner scale and raising the messenger scale.
\item At present, the allowed range of $M_2$ and $M_M$ values is more constrained by theoretical vacuum-stability constraints than by experimental data on $\BRg$ and $\amu$, though that data does mildly favor the MGM model over the standard one. However, the BNL $\amu$ experiment now in progress, and the much-awaited CLEO analysis of their $\BRg$ measurements, will dramatically constrain the MGM and allow a much more convincing discrimination between the two models. Measurements of several $\BXsll$ branching ratios and asymmetries at hadron colliders should further sharpen the b-quark side of this picture in the next few years.
\end{itemize}

\vspace{1cm}
{\bf Note added:}~
After this work was completed, new and significant theoretical 
results on radiative $b$ decay have appeared, one~\cite{ref:bsgEWcorr} 
taking into account the 
proper electroweak radiative corrections and the other~\cite{ref:kagneu} 
convincingly criticizing the current method of extracting the 
inclusive rate for $b \to s\gamma$ from the currently published CLEO data. 
The former slightly lowers the SM prediction, while the latter argues that 
theoretical uncertainties have so far been underestimated, and that only with 
more new data on the spectrum or with an improved $m_b$ value from upsilon 
spectroscopy can the uncertainties be reduced. Thus, though at present it is 
difficult to sharply compare the predicted inclusive decay rate with 
experiment, we expect that in the very near future such a comparison will be 
possible.
%%%%%%%%%%%%%%%%%%%%%%%%%%%%%%%%%%%%%%%%%%%%%%%%%%%%%%%%%%%%%%%%%%%%%%%%%%%%
\section*{Acknowledgements}
We gratefully acknowledge discussions with I. Bigi, A.L. Kagan,
T. Moroi, M. Neubert and C. Wagner.
One of us (E.G.) would like also to thank the Physics Department
of the University of Notre Dame, where most of this work was done, 
for its warm and kind hospitality.
%\newpage

\end{document}